\documentclass{article}

\usepackage{arxiv}

\usepackage[utf8]{inputenc} 
\usepackage[T1]{fontenc}    
\usepackage{hyperref}       
\usepackage{url}            
\usepackage{booktabs}       
\usepackage{amsfonts}       
\usepackage{nicefrac}       
\usepackage{microtype}      
\usepackage{lipsum}		
\usepackage{graphicx}
\usepackage{natbib}
\usepackage{doi}
\usepackage{setspace}

\usepackage{amsthm,amsmath} 
\usepackage{mathrsfs} 
\usepackage{amsmath}
\usepackage{amssymb}
\usepackage{latexsym}
\usepackage{bm}
\usepackage{color}
\usepackage{url} 
\newcommand{\choosefont}[1]{\fontfamily{#1}\selectfont}

\usepackage{mathtools} 
\usepackage{algorithmic} 
\usepackage{algorithm}
\usepackage{chngcntr} 
\usepackage[toc]{appendix} 

\theoremstyle{plain}
\newtheorem{thm}{Theorem}
\newtheorem*{thm*}{Theorem}

\theoremstyle{definition}
\newtheorem{dfn}{Definition} 

\newtheorem{lem}[thm]{Lemma} 
\newtheorem{prop}{Proposition} 

\def\argmin{\mathop{\rm argmin}}

\DeclareRobustCommand{\vect}[1]{\bm{#1}}
\title{Regularized Sparse Optimal Discriminant Clustering}

\author{ \href{https://orcid.org/0000-0000-0000-0000}{\includegraphics[scale=0.06]{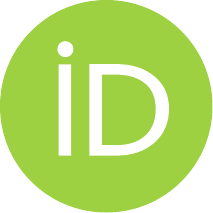}\hspace{1mm}Mayu Hiraishi}\thanks{Use footnote for providing further
		information about author (webpage, alternative
		address)---\emph{not} for acknowledging funding agencies.} \\
	Graduate School of Culture and Information Science\\
	Doshisha University\\
	Kyoto, Japan\\
	\texttt{mayumonnn@gmail.com} \\
	\And
	\href{https://orcid.org/0000-0000-0000-0000}{\includegraphics[scale=0.06]{orcid.pdf}\hspace{1mm}Kensuke Tanioka} \\
	Department of Biomedical Sciences and Informatics\\
	Doshisha University\\
	Kyoto, Japan\\
	\texttt{ktanioka@mail.doshisha.ac.jp} \\
	\And
	\href{https://orcid.org/0000-0000-0000-0000}{\includegraphics[scale=0.06]{orcid.pdf}\hspace{1mm}Hiroshi Yadohisa} \\
	Department of Culture and Information Science\\
	Doshisha University\\
	Kyoto, Japan\\
	\texttt{hyadohis@mail.doshisha.ac.jp} \\
}

\renewcommand{\shorttitle}{\textit{arXiv} Template}

\hypersetup{
pdftitle={A template for the arxiv style},
pdfsubject={q-bio.NC, q-bio.QM},
pdfauthor={David S.~Hippocampus, Elias D.~Striatum},
pdfkeywords={First keyword, Second keyword, More},
}

\begin{document}
\maketitle

\begin{abstract}

We propose a new method based on sparse optimal discriminant clustering (SODC), incorporating a penalty term into the scoring matrix based on convex clustering. With the addition of this penalty term, it is expected to improve the accuracy of cluster identification by pulling points within the same cluster closer together and points from different clusters further apart. When the estimation results are visualized, the clustering structure can be depicted more clearly. 
Moreover, we develop a novel algorithm to derive the updated formula of this scoring matrix using a majorizing function.
The scoring matrix is updated using the alternating direction method of multipliers (ADMM), which is often employed to calculate the parameters of the objective function in the convex clustering. In the proposed method, as in the conventional SODC, the scoring matrix is subject to an orthogonal constraint. Therefore, it is necessary to satisfy the orthogonal constraint on the scoring matrix while maintaining the clustering structure. 
Using a majorizing function, we adress the challenge of enforcing both orthogonal constraint and the clustering structure within the scoring matrix. 
We demonstrate numerical simulations and an application to real data to assess the performance of the proposed method.

\end{abstract}


\section{Introduction}

Dimension reduction clustering has been widely used to interpret the characteristics of large and complex data. 
It estimates a low-dimensional space for identifying clusters, allowing efficient handling while preserving important features of high-dimensional data. These methods also facilitate the interpretation of information, including visualization. 
Various dimension reduction clustering methods have been proposed ~\citep[e.g.][]{de1994k,VICHI200149,timmerman2013subspace}. 
Among these methods, we focus on optimal discriminant clustering (ODC) \citep{ODC2009} in this study.

ODC has been introduced as an unsupervised learning method based on optimal scoring for Fisher's linear discriminant analysis (LDA) \citep{osLDA}. Optimal scoring for LDA method uses class information with the scoring matrix when reducing dimensions. ODC replaces this class information matrix with an unknown scoring matrix, since it is not known in advance which cluster each subject belongs to.
The objective function of ODC has the same form as that of linear regression, and the components are defined as linear combinations of the original features. 
Additionally, the methods in the neural network field have also been proposed \citep{xie2016unsupervised}. However, due to their non-linear form, it is difficult to directly interpret which input features contribute to cluster separation. Compared with non-linear deep clustering objectives, the linearity of the objective function in ODC improves interpretability. 
Based on this method, sparse optimal discriminant clustering (SODC) was later proposed by \cite{SODC}, which incorporates a group lasso term \citep{glasso}. Both ODC and SODC describe the cluster more clearly than principal components analysis \citep{SODC}. 
However, the scoring matrix in ODC and SODC does not maintain the same structure for identifying the clusters as the optimal scoring in LDA. 
In the LDA method, the scoring matrix consists of the product of two matrices: the class information matrix and the scoring matrix, which together classify the data in reduced dimensions. 
When applying the LDA method as an unsupervised learning method, the scoring matrix in ODC and SODC is reformulated to be represented as a single matrix. 
The drawback of this single-matrix approach is that ODC and SODC do not include an independent matrix containing cluster information, which may affect the accuracy of the cluster estimation. 
From an optimization perspective, it is considered to be difficult to extend the optimal scoring in LDA method to unsupervised learning while preserving its original structures. 
In particular, it is challenging  to simultaneously satisfy the requirements of estimating a single matrix to possess both the original orthogonality constraints and the clustering structure. In addition, SODC may not produce a well-separated clustering structure when the data are
reduced to a lower dimensions and visualized the results.

Therefore, we propose a method in SODC by adding a penalty term based on convex clustering \citep{pelckmans, hocking, lindsten} to the scoring matrix, which we call regularized sparse optimal discriminant clustering (RSODC). 
With this additional term, 
the scoring matrix provides a more distinct clustering structure than the conventional SODC by drawing data points from the same cluster closer together and separating those from different clusters further apart. 
This also enables the clustering structure to be depicted more clearly when the estimated results are visualized. The clustering structure in the scoring matrix is expected to enhance the classification accuracy of the clustering. 
Unlike \cite{berends2022convex} and \cite{PCAconvex}, the model of RSODC is approximated in the reduced dimensions rather than the original dimensions. 
In addition, we develop a new algorithm using a majorizing function \citep{MMtutorial, majorizeY} to derive the updated formula of the scoring matrix with the addition of penalty term based on convex clustering. 
The scoring matrix is updated using the alternating direction method of multipliers (ADMM) \citep{boydadmm}, which is often employed to compute the parameters of the objective function for convex clustering. 
As with the conventional SODC, RSODC imposes an orthogonal constraint on the scoring matrix.
Therefore, the proposed method must satisfy the orthogonal constraint to the scoring matrix while preserving the clustering structure. 
In the ADMM process, the updated formula of the scoring matrix is derived using orthogonal Procrustes analysis \citep{schonemann}. To adress this problem, the scoring matrix needs to be expressed only in linear form, but it also contains a quadratic form. 
Therefore, we derive the majorizing function for the scoring matrix. 
This algorithm simultaneously enforces both the constraint in the scoring matrix and retains the clustering structure.

In Section $2$, we first explain studies related to the proposed method. Section $3$ presents the objective function and its algorithm of the proposed method. 
In Section $4$, we demonstrate the numerical simulations, and Section $5$ reports an application to the real gene data in . Finally, Section $6$ concludes the discussion.

\section{Related methods}
\label{sec:related}

RSODC is extended based on sparse optimal discriminant clustering (SODC). We first review related methods before presenting the proposed approach.

\subsection{Sparse optimal discriminant clustering (SODC) \label{subsec:sodc}}

Sparse optimal discriminant clustering (SODC) \citep{SODC} was developed by introducing a sparse penalty into optimal discriminant clustering (ODC) \citep{ODC2009}. ODC is a method that applies the optimal scoring for the Fisher's linear discriminant analysis (LDA) \citep{osLDA} to unsupervised learning.
Given a data matrix $\bm{X} = (\bm{x}_1, \bm{x}_2, \cdots, \bm{x}_n)^\top \in \mathbb{R}^{n \times p}$, a centering matrix $\bm{H}_n = \bm{I}_n - \frac{1}{n} \bm{1}_n \bm{1}_n^\top \in \mathbb{R}^{n\times n}$ where $\bm{1}_n = (1, 1, \cdots, 1)$ and $\bm{I}_n \in \mathbb{R}^{n \times n}$ is the identity matrix, and a known indicator matrix for class information $\bm{E} = (e_{i\ell}) \in \mathbb{R}^{n \times k} (i = 1,2,\cdots, n; \ell=1,2, \cdots k)$, where $e_{i\ell}=1$ if subject $i$ belongs to the cluster $\ell$, and $e_{i\ell}=0$ otherwise, the objective function of optimal scoring for LDA is as follows:
\begin{align*}
\min_{\bm{B}, \bm{\theta}}\frac{1}{2} \| \bm{E} \bm{\theta} - \bm{H}_n \bm{X} \bm{B} \|^2_F + \eta_2 \| \bm{B}\|_F^2
\end{align*}
s.t. $\bm{\theta}^{\top} \bm{E}^{\top}\bm{E} \bm{\theta} = \bm{I}_{k-1}$ and $ (\bm{E}  \bm{\theta})^{\top} \bm{1} = \bm{0}$,\\

\noindent
where, $\eta_2 \ (\eta_2 \geq 0)$ is a tuning parameter, $k$ is the number of classes, and 
$\| \cdot\|_F$ is the Frobenius norm.
$\bm{\theta} \in \mathbb{R}^{k \times (k-1)}$ is a scoring matrix, and its $\ell$th row indicates the score for the $\ell$th class. 
$\bm{B} = (\bm{\beta}_1, \bm{\beta}_2, \cdots, \bm{\beta}_p)^\top \in \mathbb{R}^{p \times (k-1)}$ is the weight for each variable used to interpret $\bm{X}\bm{B}$, the coordinates in the low-dimensional space, which are estimated in the form of a regression with $\bm{E} \bm{\theta}$ as the objective variable.
Each data $\bm{x}_i$ is assumed to belong to one group. 

In unsupervised learning, \cite{ODC2009} proposed optimal discriminant clustering (ODC) based on this optimal scoring for LDA. Class information is already known in the supervised learning, whereas in unsupervised learning method, no information is given a priori about which cluster a subject belongs to. 
Therefore, ODC modified the class information variables $\bm{E} \bm{\theta}$ in optimal scoring for LDA to $\bm{Y}^{\dagger} = (\bm{y}_1^\dagger,\bm{y}_2^\dagger, \cdots, \bm{y}_n^\dagger )^\top \in \mathbb{R}^{n \times (k-1)}$, treated as unknown scoring matrix. 
$\bm{Y}^\dagger$ is regarded as containing cluster information. 
This objective function is 

\begin{align}
\min_{\bm{B}, \bm{Y}^{\dagger}}\frac{1}{2} \| \bm{Y}^{\dagger} - \bm{H}_n \bm{X} \bm{B} \|^2_F + \eta_2 \| \bm{B}\|_F^2
\label{odc}
\end{align}
s.t. $\bm{Y}^{\dagger \top} \bm{Y}^{\dagger} = \bm{I}_{k-1}$ and $ \bm{Y}^{\dagger \top} \bm{1}= \bm{0}$.\\

ODC imposes the same constraints on $\bm{Y}^{\dagger}$ as on $\bm{E} \bm{\theta}$ in the optimal scoring for LDA. 
Based on Eq. (\ref{odc}), SODC was proposed by adding a sparse penalty, and its objective function is

\begin{align}
\min_{\bm{B}, \bm{Y}^{\dagger}}\frac{1}{2} \| \bm{Y}^{\dagger} - \bm{H}_n \bm{X} \bm{B} \|^2_F + \eta_2 \| \bm{B}\|_F^2 + \eta_1 \sum_{j=1}^p \| \bm{\beta}_j \|_2
\label{sodc}
\end{align}
s.t. $\bm{Y}^{\dagger \top} \bm{Y}^{\dagger} = \bm{I}_{(k-1)}$ and $\bm{Y}^{\dagger \top} \bm{1}= \bm{0}$.\\

Here, $\eta_1 \ (\eta_1\geq 0)$ and $\eta_2 \ (\eta_2 \geq 0)$ are tuning parameters, and $\|\cdot \|_2$ is the Euclidean norm. SODC includes a group lasso penalty \citep{glasso} for variable selection. 
Eq. (\ref{sodc}) is used to obtain the final solution by alternately updating the parameters: $\bm{Y}^\dagger$ is updated by singular value decomposition, and $\bm{B}$ is updated in a manner similar to the group lasso.

\subsection{Convex clustering \label{subsec:convex}}

Convex clustering \citep{pelckmans, hocking, lindsten} is a method for dividing data points into clusters using convex optimization to achieve stable clustering. It detects the clustering structure by finding parameters that minimize an objective function, which includes a penalty term based on the distances between data points.

Given a data matrix $\bm{X} \in \mathbb{R}^{n \times p}$, the objective function of convex clustering is defined as follows:

\begin{align}
\min_{\bm{M} \in \mathbb{R}^{n \times p}}\frac{1}{2}  \| \bm{X} - \bm{M} \|_2^2
+ \gamma \sum_{i < j} \alpha_{i, j} \| \bm{m}_{i} - \bm{m}_{j} \|_2
\label{convex}
\end{align}

where $\gamma \ (\gamma \geq 0)$ is a tuning parameter. 
$\bm{M} = (\bm{m}_1, \bm{m}_2, \cdots, \bm{m}_n)\in \mathbb{R}^{n \times p}$ represents the cluster centroids, and $\bm{m}_{i} \in \mathbb{R}^p$ denotes the $i$th row of $\bm{M}$. 
Through penalty terms, 
the cluster centroids belonging to the same cluster are estimated as the same value. 
Consequently, if $\bm{\hat{m}}_i =\bm{\hat{m}}_j$, the $i$th and $j$th data points are assigned the same cluster. 
$\alpha_{i, j} \ (\alpha_{i, j}\geq 0)$ is the weight. Several methods have been proposed to compute this weight \citep{hocking, chiadmm}. In this study, we adopt the following formulation of \cite{chiadmm}:

\begin{align}
\alpha_{i, j}= \iota^{\delta}_{i ,j} \exp (-\tau\|\bm{x}_{i}- \bm{x}_{j } \|^2_2)
\label{alpha_weight}
\end{align}

where $\iota^{\delta}_{i\cdot ,j}$ returns $1$ if $j$ is among $i$'s $\delta$ nearest neighbors or if $i$ is among $j$'s $\delta$ nearest neighbors, returns $0$ otherwise. 
$\tau \ (\tau \geq 0)$ is a tuning parameter. 
For the second term of Eq. (\ref{convex}), convex clustering with $L_1$, $L_2$, and $L_\infty$ penalties has been proposed \citep{boydadmm}. In this study, we focus on the $L_2$ norm in this study. To solve this optimization problem, we apply the alternating direction method of multipliers (ADMM) \citep{boydadmm, chiadmm}. For the implementation of ADMM, Eq. (\ref{convex}) can be rewritten in the equivalent form:

\begin{align}
\min_{\bm{M}, \bm{v}_l }\frac{1}{2}  \| \bm{X} - \bm{M} \|_2^2
+ \gamma \sum_{l \in \varepsilon} \alpha_l \| \bm{v}_l \|_2,
\label{convex2}
\end{align}

s.t. $(\bm{m}_i- \bm{m}_j) - \bm{v}_l = \bm{0}$. \\

Here, $l = (i,j)$ is a pair of subjects, and $\varepsilon = \{ l = (i,j): \alpha_l > 0, \ i<j, \ i,j=1, 2, \cdots, n \}$. $\bm{v}_l$ $\in \mathbb{R}^{(n(n-1)/2) \times p}$ represents the difference between $\bm{m}_i$ and $\bm{m}_j$. 
By introducing the penalty term through an auxiliary variable, ADMM separates the objective function into two components, allowing parameters to be updated efficiently.

\section{Regularized Sparse Optimal Discriminant Clustering (RSODC)}
\label{sec:prop}

We introduce the proposed method, termed Regularized Sparse Optimal Discriminant Clustering (RSODC). First, we present the objective function of RSODC in \ref{subsec:obj}. 
Next, in \ref{subsec:algorithm}, we explain the overall algorithm. 
Finally, we provide the updated formula for each parameter in \ref{subsec:updB} to \ref{subsec:algY}.

\subsection{Optimization problem of RSODC}
\label{subsec:obj}

We propose a new method for SODC by adding a regularization term from convex clustering  to provide more discriminative optimal scoring of SODC. 
By incorporating this regularization term, ADMM algorithm is employed. To facilitate the solution, the proposed method must be expressed in the form of an augmented Lagrangian function. In this subsection, we first define the optimization problem of the proposed method and then describe the transformation to its augmented Lagrangian function. 

Given data $\bm{X}$ and centering matrix $\bm{H}_n$, the optimization problem is defined as

\begin{align}
  \min_{\bm{B},\bm{Y}^{\dagger}} 
  \frac{1}{2}\|\bm{Y}^{\dagger} -\bm{H}_n\bm{X}\bm{B}\|^2_F + \eta_2 \|\bm{B}\|^2_F+\eta_1 \sum_{j=1}^p\|\bm{\beta}_j\|_2 +\gamma \sum_{i <j} \alpha_{i, j} \| \bm{y}^{\dagger}_i - \bm{y}^{\dagger}_j\|_2, 
\label{prop}
\end{align}

s.t. $\bm{Y}^{\dagger \top} \bm{Y}^{\dagger} = \bm{I}_{k-1}$ and $\bm{Y}^{\dagger \top} \bm{1}=\bm{0}$. \\

Here,  $\eta_1 (\eta_1\geq 0), \eta_2 (\eta_2\geq 0)$ and $\gamma (\gamma\geq 0)$ are tuning parameters. 
$\bm{B} = (\bm{\beta}_1, \bm{\beta}_2, \cdots, \bm{\beta}_p)^\top = \big(\bm{\beta}_{(1)}, \bm{\beta}_{(2)},\cdots, \bm{\beta}_{(k-1)}\big) \in \mathbb{R}^{p \times (k-1)}$ is the weight for each variable. 
The first three terms of Eq. (\ref{prop}) are identical to those in SODC.  
We add a penalty term in convex clustering \citep{hocking} to SODC as a penalty term to $\bm{Y}^{\dagger} \in \mathbb{R}^{n \times (k-1)}$, corresponding to the fourth term of Eq. (\ref{prop}). 
This additional term encourages similar subjects to be placed closer together. 
The constraints on $\bm{Y}^\dagger$ are same as in the conventional SODC. 
$\alpha_{i, j} = \alpha_l(\alpha_l\geq 0)$ is a weight, which is calculated using Eq. (\ref{alpha_weight}). 
A larger $\alpha_{i ,j}$ brings the $L_2$ norm of $\bm{y}_{i}^\dagger - \bm{y}_{j}^\dagger$ closer to $\bm{0}$. 
To solve Eq. (\ref{prop}), we apply the ADMM algorithm commonly used for solving convex clustering \citep{chiadmm}. 
To apply ADMM, the augmented Lagrangian function of Eq. (\ref{prop}) needs to be solved. 
For this, Eq. (\ref{prop}) is rewritten as 

\begin{align}
  \min_{\bm{B},\bm{Y}, \bm{V}, \bm{\Lambda}} 
  \frac{1}{2}\|\bm{Y}-\bm{H}_n\bm{X}\bm{B}\|^2_F + \eta_2 \|\bm{B}\|^2_F+\eta_1 \sum_{j=1}^p\|\bm{\beta}_j\|_2 +\gamma \sum_{l \in \varepsilon} \alpha_l \|\bm{v}_l\|_2, 
\label{prop_v}
\end{align}
s.t. 
\begin{align}
\bm{y}_{i} - \bm{y}_{j} &=\bm{v}_l, \label{constr1}\\
\bm{Y}^\top\bm{Y} &= \bm{I}_{k-1}, {\rm and} \label{constr2}\\
\bm{Y}^\top \bm{1} &=\bm{0}. \label{constr3}
\end{align}

$\bm{V} = (\bm{v}_1, \bm{v}_2, \cdots, \bm{v}_{|\varepsilon|})^\top$, where $|\cdot|$ denotes the cardinality of a set, is represents the pairwise differences between two cluster centroids. 
In Eq. (\ref{prop_v}), $\bm{v}_l$ is substituted as a penalty term corresponding to the constraints Eq. (\ref{constr1}). 
In solving the optimization problem of Eq. (\ref{prop}),
the optimal scoring $\bm{Y}^\dagger$ must be described as two distinct parameters: $\bm{Y} = (\bm{y}_{(1)}, \bm{y}_{(2)},\cdots, \bm{y}_{(k-1)}) \in \mathbb{R}^{n \times (k-1)} \ (\bm{y}_{(j)} \in \mathbb{R}^n, j = 1, 2, \cdots, k-1)$ and $\bm{V}$ for estimation. Therefore, $\bm{Y}$ is denoted separately from $\bm{Y}^\dagger$ in Eq. (\ref{prop}) to distinguish between them.  
Using the ADMM algorithm, the problem of estimating $\bm{Y}^\dagger$ is treated as one solved by decomposing it into $\bm{Y}$ and $\bm{V}$ in Eq. (\ref{prop_v}). 
Then, solving Eq. (\ref{prop_v}) is equivalent to minimizing the augmented Lagrangian function, which can be defined as

\begin{align}
  \min_{\bm{B},\bm{Y}, \bm{V}, \bm{\Lambda}} \frac{1}{2}\|\bm{Y}-\bm{H}_n\bm{X}\bm{B}\|^2_F 
  &+ \eta_2 \|\bm{B}\|^2_F
  + \eta_1 \sum_{j=1}^p\|\bm{\beta}_j\|_2 
 +\gamma \sum_{l \in \varepsilon} \alpha_l \|\bm{v}_l\|_2 \nonumber\\
 &+ \sum_{l\in\varepsilon}\bm{\lambda}_l^\top(\bm{v}_l- \bm{y}_{i} + \bm{y}_{j} ) 
 +\frac{\rho}{2}\sum_{l\in\varepsilon}\|\bm{v}_l- \bm{y}_{i} + \bm{y}_{j} \|_2^2,
\label{prop_lag}
\end{align}

where $\rho (\rho>0)$ is a tuning parameter, and $\bm{\Lambda} =(\bm{\lambda}_1, \bm{\lambda}_2, \cdots, \bm{\lambda}_{|\varepsilon|})^\top$ is the Lagrangian multiplier.

\subsection{Algorithm}
\label{subsec:algorithm}

This subsection presents the general framework of the algorithm of RSODC is presented. {\bf STEP 2.} contains further update steps associated with the inclusion of the penalty term.

As in SODC, the alternating least squares (ALS) method is used to update $\bm{B}$ and $\bm{Y}^\dagger$. For updating $\bm{Y}^{\dagger}$, we adopt the alternating direction method of multipliers (ADMM) \citep{boydadmm, chiadmm}. 
The overall algorithm is the following steps:\\
oindent
{\bf Algorithm of RSODC}

\noindent
{\bf STEP 0.} Set initial value of $\bm{B}, \bm{Y}^{\dagger}, \bm{V}$, and $\bm{\Lambda}$.

\noindent
{\bf STEP 1.} Update $\bm{B}$ under given $\bm{Y}^{\dagger}$.

\noindent
{\bf STEP 2.} Update $\bm{Y}^{\dagger}$ under given $\bm{B}$ by using ADMM algorithm.

\noindent
\quad {\bf STEP 2-1.} Update $\bm{Y}$ under given $\bm{V}$ and $\bm{\Lambda}$.

\noindent
\quad {\bf STEP 2-2.} Update $\bm{V}$ under given $\bm{Y}$ and $\bm{\Lambda}$.

\noindent
\quad {\bf STEP 2-3.} Update $\bm{\Lambda}$ under given $\bm{Y}$ and $\bm{V}$.

\noindent
\quad {\bf STEP 2-4.} Repeat {\bf STEP 2-1.} to {\bf STEP 2-3.} until the value of the objective function converges.

\noindent
{\bf STEP 3.} Repeat {\bf STEP 1.} and {\bf STEP 2.} until the value of the objective function converges.

\noindent
{\bf STEP 4.} Calculate $\bm{X}^\dagger = (\bm{x}^\dagger_1, \bm{x}^\dagger_2, \cdots, \bm{x}^\dagger_n)^\top = \bm{H}_n \bm{X} \hat{\bm{B}}$.

\noindent
{\bf STEP 5.} Apply {\it k}-means to $\bm{X}^\dagger$.

\noindent
{\bf STEP 6.} Obtain the clustering result of $\bm{X}^\dagger$.\\

In {\bf STEP 2-1.}, the proposed method must satisfy the constraints of SODC, in Eq. (\ref{constr2}) and Eq. (\ref{constr3}), as well as the constraint for the penalty term on $\bm{Y}$ in Eq. (\ref{constr1}) simultaneously. However, it is difficult to solve this problem by directly applying the updated procedure of the conventional SODC.  Therefore, we propose a novel algorithm for $\bm{Y}$ by deriving the majorizing function. 
In {\bf STEP 2-5.}, we adopt {\it k}-means following the same procedure as the conventional SODC.

The updated formula for $\bm{B}$ will be explained in \ref{subsec:updB}. Then, the deriving the majorization function for the updated formula of $\bm{Y}$ will be presented in \ref{subsec:maj_socdc}. After that, the update procedure for $\bm{Y}^\dagger$ will be described in \ref{subsec:algY}.\\

The hyperparameters $\eta_1, \eta_2, \gamma$, and $\rho$ are determined using a modified version of cross-validation based on the kappa coefficient \citep{cohen1960coefficient} proposed by \citep{SODC, Sun2013} with the idea of clustering stability. 
To simplify the calculation, $\eta_2$ is set as $0$ in this study.

\subsection{Update formula of $\vect{B}$ \label{subsec:updB}} 

In this subsection, we explain the updated formula for $\bm{B}$. This uses the coordinate descent algorithm \citep{coorde} for the group lasso update procedure.

First, to update in the form of linear regression, the terms involving $\bm{B}$ are transposed in vector form. In this study, we set $\eta_2=0$. For the terms related to $\bm{B}$ in Eq. (\ref{prop}),

\begin{align}
&\|\bm{Y}-\bm{H}_n\bm{X}\bm{B}\|^2_F + \eta_2 \|\bm{B}\|^2_F
+\eta_1 \sum_{j=1}^p\|\bm{\beta}_j\|_2 \nonumber\\
\Longleftrightarrow & \|\bm{y}^*-\bm{Z}\bm{\beta}^*\|^2_F +\eta_1 \sum_{j=1}^p\|\bm{\beta}_j\|_2,
\label{vecto}
\end{align}
where 
${\it vec}(\bm{B})= \bm{\beta}^* = \big(\bm{\beta}_{(1)}^\top, \bm{\beta}_{(2)}^\top,\cdots, \bm{\beta}_{(k-1)}^\top \big)^\top \in \mathbb{R}^{p(k-1)\times 1}$. 

Here,

\begin{align*}
{\it vec}(\bm{Y})=& \big(\bm{y}_{(1)}^\top, \bm{y}_{(2)}^\top,\cdots, \bm{y}_{(k-1)}^\top \big)^\top \in \mathbb{R}^{n(k-1)\times 1},\\
\bm{Z}^* =&(\bm{z}^*_{(1)},\bm{z}^*_{(2)}, \cdots, \bm{z}^*_{(p)}), {\rm where} \ \bm{H}_n\bm{X}=\bm{z}^*.
\end{align*}

Then, we set $\bm{y}^*$ and $\bm{Z}$ as follows:

\begin{align*}
\bm{y}^* = \left(
\begin{array}{cc}
{\it vec}(\bm{Y}) \\
\bm{0}_{p(k-1)}
\end{array} \right), \
\bm{Z}= \left(
\begin{array}{cc}
\mathrm{bdiag}(\bm{Z}^*)\\
\sqrt{\eta_2} \bm{I}_{p(k-1)}
\end{array}\right).
\end{align*}

This algorithm is updated for every $\bm{\beta}_j  \ (j=1, 2, \cdots, p)$.
Here, $\bm{Z}_j \in \mathbb{R}^{(n+p)(k-1) \times (k-1)}$ is the submatrix of $\bm{Z}$ corresponding to the $j$th variables of $\bm{X}$.

\begin{prop}\label{prp2}

Given $\bm{Y}, \bm{Z}, \eta_1 \ (\eta_1 \geq 0), \eta_2 \ (\eta_2 \geq 0),\nu \ (\nu>0)$, 
the updated formula for $\bm{B}$ is derived as follows:

\begin{align}
\bm{\beta}_j \leftarrow
\begin{cases}
\bm{\phi} \bigg(1-\frac{\nu\eta_1}{\|\bm{\phi} \|_2} \bigg), \  &(\|\bm{\phi} \|_2 > \nu \eta_1)\\
\bm{0}, \ &(\| \bm{\phi}\|_2 \leq \nu\eta_1)
\end{cases}
\label{upd_beta}
\end{align}

where $\bm{\phi} = \bm{\beta}_j +\nu \bm{Z}_j^{\top}(\bm{r}_j-\bm{Z}_j\bm{\beta}_j)$,
$\bm{r}_j = \bm{y}^* -  \sum_{o \neq j}\bm{Z}_o \bm{\beta}_o$,
and $\nu$ is a threshold parameter for updating. 

\begin{proof} 

For any $j$, we obtain the following equation from Eq. (\ref{vecto}) :

\begin{align}
\Big\|\bm{y}^*- \bigg(\bm{Z}_j\bm{\beta}_j + \sum_{o\neq j} \bm{Z}_o \bm{\beta}_o \bigg) \Big\|^2_F 
+ \eta_1 \|\bm{\beta}_j\|_2.
\label{group1}
\end{align}

When $\bm{r}_j = \bm{y}^* -  \sum_{o \neq j}\bm{Z}_o \bm{\beta}_o$, Eq. (\ref{group1}) can be rewritten as

\begin{align}
\Big\|\bm{r}_j - \bm{Z}_j\bm{\beta}_j \Big\|^2_F + \eta_1 \|\bm{\beta}_j\|_2.
\label{group11}
\end{align}

Since Eq. (\ref{group11}) has the same equation as the conventional group lasso, it can be solved using the proximal gradient method for group lasso. 
For a given $\bm{\beta}_j$, Eq. (\ref{group11}) can be divided into the differentiable term and the non-differentiable term.

\begin{align}
 \begin{cases}
f(\bm{\beta}_j) = \big\|\bm{r}_j -\bm{Z}_j\bm{\beta}_j \big\|^2_F \\
g(\bm{\beta}_j) = \eta_1 \|\bm{\beta}_j\|_2
\end{cases}
\label{prox_pro}
\end{align}

First, $f(\bm{\beta}_j)$ differentiated with respect to $\bm{\beta}_j$ can be obtained as $-\bm{Z}^{\top}_j\bm{r}_j + \bm{Z}_j^{\top} \bm{Z}_j\bm{\beta}_j$. The gradient for $f(\bm{\beta}_j)$ is then expressed as follows, which we denote by $\bm{\phi}$.  

\begin{align}
\bm{\phi}
\equiv & \bm{\beta}_j - \nu( -\bm{Z}^{\top}_j\bm{r}_j+ \bm{Z}_j^{\top}\bm{Z}_j\bm{\beta}_j)\nonumber\\
=&\bm{\beta}_j + \nu \bm{Z}_j^{\top}
(\bm{r}_j - \bm{Z}_j\bm{\beta}_j )
\label{grad}
\end{align}

Then, using Eq. (\ref{grad}), the proximal operator of $g(\bm{\beta}_j)$ is defined as follows:

\begin{align}
{\bf prox}_{\nu g}(\bm{\phi})
= \argmin_{\bm{\beta}_j} \bigg( \nu \eta_1 \| \bm{\beta}_j \|_2 +\frac{1}{2} \| \bm{\beta}_j- \bm{\phi} \|_2^2\bigg).
\label{prox_w}
\end{align}

When $\bm{\phi} \neq \bm{0}$, Eq. (\ref{prox_w}) is differentiated with respect to $\bm{\beta}_j$, and set equal to $\bm{0}$.

\begin{align*}
\nu \eta_1 \frac{\bm{\beta}_j}{\|\bm{\beta}_j \|_2}
+(\bm{\beta}_j - \bm{\phi})   =& \bm{0}\\
\Longleftrightarrow \hspace{40mm} \bm{\beta}_j \bigg(
1+  \nu \eta_1 \frac{1}{\|\bm{\beta}_j \|_2} 
\bigg) =&  \bm{\phi} \tag{i} \\
\Longleftrightarrow \hspace{6.6mm}
\bigg(
1+  \nu \eta_1 \frac{1}{\|\bm{\beta}_j \|_2} 
\bigg)^\top \bm{\beta}_j^\top \bm{\beta}_j
\bigg(
1+  \nu \eta_1 \frac{1}{\|\bm{\beta}_j \|_2} 
\bigg) =& \bm{\phi}^\top \bm{\phi}\\
\Longleftrightarrow \hspace{33.5mm}
\bigg(
1+  \nu \eta_1 \frac{1}{\|\bm{\beta}_j \|_2} 
\bigg)^2 \| \bm{\beta}_j \|_2^2 = & \| \bm{\phi} \|_2^2 \\
\Longleftrightarrow \hspace{35mm}
\bigg(
1+  \nu \eta_1 \frac{1}{\|\bm{\beta}_j \|_2} 
\bigg) \| \bm{\beta}_j \|_2 = & \| \bm{\phi} \|_2 \\
\Longleftrightarrow \hspace{52mm}
\| \bm{\beta}_j \|_2 + \nu \eta_1= & \|\bm{\phi}\|_2\\
\Longleftrightarrow \hspace{61.5mm}
\| \bm{\beta}_j \|_2 =& \|\bm{\phi}\|_2 - \nu \eta_1 \tag{ii}
\end{align*}

Therefore, substitute Eq.(ii) into Eq.(i).
\begin{align*}
\bm{\beta}_j \bigg(
1+  \nu \eta_1 \frac{1}{\|\bm{\phi}\|_2 - \nu \eta_1 }
\bigg) =&  \bm{\phi}\\
\Longleftrightarrow \hspace{5mm}
\bm{\beta}_j \bigg(
\frac{\|\bm{\phi}\|_2 - \nu \eta_1 + \nu \eta_1}{\|\bm{\phi}\|_2 - \nu \eta_1 }
\bigg) =&  \bm{\phi}\\
\Longleftrightarrow \hspace{37.5mm}
\bm{\beta}_j =&  \bm{\phi} \bigg(
1- \frac{\nu \eta_1 }{\|\bm{\phi}\|_2 }
\bigg) \\
\end{align*}
\end{proof}
\end{prop}

\subsection{Deriving the majorizing function for updated formula of $\vect{Y}$ \label{subsec:maj_socdc}}

Before presenting the updated formula for $\bm{Y}^\dagger$, we introduce a novel algorithm to derive the updated formula for $\bm{Y}$ by using a majorizing function, which corresponds to {\bf STEP 2-1.} in this subsection. 
$\bm{Y}^\dagger$ is updated by alternately minimizing with respect to $\bm{Y}$, $\bm{V}^\dagger$, and $\bm{\Lambda}$ using the ADMM algorithm. 
We show that the updated formula of $\bm{Y}$ is equivalent to the orthogonal Procrustes problem \citep{schonemann} for a certain matrix $\bm{D} \in \mathbb{R}^{n \times (k-1)}$. 
Here, we briefly explain the orthogonal Procrustes problem.

For $\bm{D}$, the optimization problem of the orthogonal Procrustes analysis is

\begin{align}
\min_{Y}\|\bm{Y}-\bm{D} \|^2_F,
\label{proc}
\end{align}

s.t. $\bm{Y}^\top \bm{Y}=\bm{I}$.\\

For this problem, the solution $\bm{Y}$ that minimizes Eq. (\ref{proc}) is given by $\bm{Y} = \bm{L} \bm{R}^\top$, where the singular value decomposition is $\bm{D} = \bm{L} \bm{\Sigma} \bm{R}^\top$. Here, $\bm{L}$ is the left singular matrix, $\bm{R}$ is the right singular matrix, and $\bm{\Sigma}$ is the singular value matrix. Expanding Eq. (\ref{proc}), the terms related to $\bm{Y}$ are 

\begin{align}
\|\bm{Y}-\bm{D} \|^2_F = \mathrm{tr}(\bm{Y}^\top\bm{Y}) - 2\mathrm{tr}(\bm{Y}^\top \bm{D}) + \mathrm{const},
\label{procru}
\end{align}

where $\mathrm{const}$ is a constant value.

Eq. (\ref{procru}) can be solved as follows:

\begin{align*}
\bm{Y} \leftarrow \bm{L}\bm{R}^\top,
\end{align*}

where $\bm{L}$ and $\bm{R}$ are obtained from the singular value decomposition $\bm{D} = \bm{L} \bm{\Sigma} \bm{R}^\top$.

We now show the procedure for deriving the updated formula for $\bm{Y}$ from the proposed method using the orthogonal Procrustes analysis, when the other parameters are given.
The terms containing $\bm{Y}$ in Eq. (\ref{prop_lag}) can be rewritten as follows:

\begin{align}
& \frac{1}{2}\|\bm{Y}-\bm{H}_n\bm{X}\bm{B} \|^2_F +\sum_{l\in\varepsilon}\bm{\lambda}_l^\top(\bm{v}_l-\bm{y}_{i} + \bm{y}_{j})
 +\frac{\rho}{2}\sum_{l\in\varepsilon}\|\bm{v}_l- \bm{y}_{i} + \bm{y}_{j}\|_2^2 \nonumber\\
= & \frac{1}{2} \|\bm{Y}-\bm{W} \|^2_F +\sum_{l\in\varepsilon}\bm{\lambda}_l^\top(\bm{v}_l- \bm{Y}^\top \bm{g}_l)
 +\frac{\rho}{2}\sum_{l\in\varepsilon}\|\bm{v}_l- \bm{Y}^\top \bm{g}_l\|_2^2.
\label{okikae_y}
\end{align}

\noindent
where $\bm{H}_n\bm{X} \bm{B} = \bm{W}$ and $\bm{y}_{i} -\bm{y}_{j} = \bm{Y}^\top \bm{g}_l$.
Here, $\bm{g}_l = (g_{1l}, g_{2l}, \cdots, g_{nl})^\top$, $g_{il}= 1, g_{jl}=-1$, and $g_{m^\dagger l}=0 \ (m^\dagger \neq i, j)$. 

From Eq. (\ref{okikae_y}), we have

\begin{align}
&  \frac{1}{2} \|\bm{Y}- \bm{W} \|^2_F +\sum_{l\in\varepsilon}\bm{\lambda}_l^\top(\bm{v}_l- \bm{Y}^\top \bm{g}_l)
 +\frac{\rho}{2}\sum_{l\in\varepsilon}\|\bm{v}_l- \bm{Y}^\top \bm{g}_l\|_2^2 \nonumber\\
=&  -  \mathrm{tr} (\bm{Y}^\top \bm{W})  - \mathrm{tr} \Big(\sum_{l \in \varepsilon}\bm{\lambda}_l^\top \bm{Y}^\top \bm{g}_l \Big)
- \rho \mathrm{tr} \Big(\sum_{l \in \varepsilon}\bm{Y}^\top \bm{g}_l \bm{v}^{\top}_l \Big)
+ \frac{\rho}{2} \mathrm{tr} \Big(\sum_{l \in \varepsilon}\bm{Y}^\top \bm{g}_l\bm{g}_l^\top \bm{Y} \Big) + \mathrm{const},
\label{tochu_y}
\end{align}

\noindent
where $\mathrm{const}$ is a term not related to $\bm{Y}$.
The terms related to $\bm{Y}$ in Eq. (\ref{tochu_y}) can be calculated as 

\begin{align}
& - \mathrm{tr}(\bm{Y}^\top \bm{W})
- \mathrm{tr} \Big(\bm{Y}^\top \sum_{l \in \varepsilon}\bm{g}_l\bm{\lambda}_l^\top \Big)
- \rho \mathrm{tr} \Big(\bm{Y}^\top \sum_{l \in \varepsilon}\bm{g}_l \bm{v}_l \Big)
+ \frac{\rho}{2}\mathrm{tr} \Big(\bm{Y}^\top \sum_{l \in \varepsilon}\bm{g}_l\bm{g}_l^\top \bm{Y} \Big) \nonumber\\
=&  - \mathrm{tr} \bigg( \bm{Y}^\top \Big(\bm{W}
+\sum_{l \in \varepsilon}\bm{g}_l\bm{\lambda}_l^\top
+ \rho \sum_{l \in \varepsilon}\bm{g}_l \bm{v}_l^{\top} \Big)\bigg)
+ \frac{\rho}{2}\mathrm{tr}\Big(\bm{Y}^\top \sum_{l \in \varepsilon}\bm{g}_l\bm{g}_l^\top \bm{Y}\Big).
\label{tenkai_y}
\end{align}

The first, second, and third terms of Eq. (\ref{tenkai_y}) can be expressed in the form of linear function of $\bm{Y}$ such as $\mathrm{tr}(\bm{Y}^\top \bm{D})$, where $\bm{D}$ is a certain matrix. 
However, the fourth term of Eq. (\ref{tenkai_y}) is quadratic form, not linear function. It is difficult to solve this orthogonal Procrustes analysis, unless this term is transformed into a linear form. To adress this problem, RSODC employs a majorizing function \citep{MMtutorial, majorizeY}.

A Majorizing function is a function defined in the majorization-minimization (MM) algorithm. 
The MM algorithm is an algorithm that can optimize the objective function without requiring hyperparameters. 
Let $\bm{\theta} \in \mathbb{R}^p$ denote the parameter of the real-valued objective function of interest $f:\mathbb{R}^p \mapsto \mathbb{R}$. $\bm{\theta}^{(t)}$ represents the estimates of $\bm{\theta}$ at the $t$th step in the algorithm.
$g: \mathbb{R}^p \mapsto \mathbb{R}$ depending on $\bm{\theta}^{(t)}$ is also a real-valued function such that the updated formula can be easily derived. This function $g(\bm{\theta} | \bm{\theta}^{(t)})$ is then defined as the majorizing function of $f(\bm{\theta})$ in $\bm{\theta}^{(t)}$, when the following two conditions are satisfied:

\begin{align*}
g(\bm{\theta}|\bm{\theta}^{(t)}) & \geq f(\bm{\theta}) \ {\rm for \ all} \ \bm{\theta},\\
g(\bm{\theta}^{(t)}|\bm{\theta}^{(t)}) &= f(\bm{\theta}^{(t)}).\\
\end{align*}

We derive the majorizing function for Eq. (\ref{tenkai_y}) in the same manner as in \cite{majorizeY} and \cite{MMconvex}.

\begin{lem}\label{lem1}

Given $\bm{Y}, \bm{C}, \bm{Q},$ and $\omega$, the following inequality holds:
\begin{align}
\mathrm{tr}(\bm{Y}^\top \bm{C}\bm{Y}) \leq 2 \omega - 2\mathrm{tr}(\bm{Y}^\top (\omega\bm{I}-\bm{C})\bm{Q} - \mathrm{tr}(\bm{Q}^\top \bm{C}\bm{Q}),
\label{maj_ymat}
\end{align}

where $\bm{C} = \frac{\rho}{2}\sum_{l \in \varepsilon}\bm{g}_l\bm{g}_l^\top$, $\bm{Q}$ is $\bm{Y}$ from the previous step, $\omega \ (\omega >0)$ is the largest eigenvalue of $\bm{C}$, and $\bm{C} - \omega \bm{I}$ is negative semidefinite.

\noindent
\begin{proof}

For any $\bm{y}_{\varrho}$, in the case that $\bm{y}_{\varrho}^\top \bm{y}_{\varrho}=\bm{q}_{\varrho}^\top \bm{q}_{\varrho}=1$, the following inequality holds:

\begin{align*}
\bm{y}_{\varrho}^\top \bm{C}\bm{y}_{\varrho} \leq 2 \omega - 2\bm{y}_{\varrho}^\top (\omega\bm{I}-\bm{C})\bm{q}_{\varrho} - \bm{q}_{\varrho}^\top \bm{C}\bm{q}_{\varrho}.
\end{align*}

Since it holds for any $\bm{y}_{o^\dagger}$, the following inequality holds:

\begin{align}
\sum_{\varrho=1}^d\bm{y}_{\varrho}^\top \bm{C}\bm{y}_{\varrho}
\leq \sum_{\varrho=1}^d(2\omega-2\bm{y}_{\varrho}^\top (\omega\bm{I}-\bm{C})\bm{q}_{\varrho} -\bm{q}_{\varrho}^\top \bm{C}\bm{q}_{\varrho}).
\label{maj_y}
\end{align}

Here, $d=k-1$. 
For the left-hand side of Eq. (\ref{maj_y}), the matrix notation can be written as $\mathrm{tr}(\bm{Y}^\top \bm{C}\bm{Y}) = \sum_{\varrho=1}^d\bm{y}_{\varrho}^\top \bm{C}\bm{y}_{\varrho}$.

Using Eq. (\ref{maj_y}), this can be expressed as

\begin{align*}
\sum_{\varrho=1}^d\bm{y}_{\varrho}^\top \bm{C}\bm{y}_{\varrho}
&\leq \sum_{\varrho=1}^d(2\omega-2\bm{y}_{\varrho}^\top (\omega\bm{I}-\bm{C})\bm{q}_{\varrho} -\bm{q}_{\varrho}^\top \bm{C}\bm{q}_{\varrho})\nonumber\\
\Longleftrightarrow \quad \mathrm{tr}(\bm{Y}^\top \bm{C}\bm{Y}) &\leq 2 \omega - 2\mathrm{tr}(\bm{Y}^\top (\omega\bm{I}-\bm{C})\bm{Q} - \mathrm{tr}(\bm{Q}^\top \bm{C}\bm{Q}),
\end{align*}

where $\bm{Y}^\top \bm{Y}=\bm{Q}^\top \bm{Q}=\bm{I}$.\\
\end{proof}
\end{lem}

\begin{lem}\label{lem2}

The following inequality is satisfied:

\begin{align}
&\frac{1}{2}\|\bm{Y}-\bm{H}_n\bm{X}\bm{B} \|^2_F +\sum_{l\in\varepsilon}\bm{\lambda}_l^\top(\bm{v}_l-\bm{y}_{i} + \bm{y}_{j})
 +\frac{\rho}{2}\sum_{l\in\varepsilon}\|\bm{v}_l- \bm{y}_{i} + \bm{y}_{j}\|_2^2 \nonumber\\
\leq & - \mathrm{tr}\bm{Y}^\top \Big(\bm{W}
+\sum_{l \in \varepsilon}\bm{g}_l\bm{\lambda}_l^\top
+ \rho \sum_{l \in \varepsilon}\bm{g}_l \bm{v}_l^{\top} 
+ 2(\omega\bm{I}-\bm{C})\bm{Q}\Big) + \mathrm{const}, 
\label{lemma2} 
\end{align}

where $\mathrm{const}$ denotes terms not relevant to $\bm{Y}$.

\begin{proof}
From Eq. (\ref{maj_ymat}) in ${\bf Lemma \ \ref{lem1}}$, the majorizing function of Eq. (\ref{tenkai_y}) can be derived as follows:

\begin{align}
&\frac{1}{2} \|\bm{Y}-\bm{W} \|^2_F +\sum_{l\in\varepsilon}\bm{\lambda}_l^\top(\bm{v}_l- \bm{Y}^\top \bm{g}_l)
 +\frac{\rho}{2}\sum_{l\in\varepsilon}\|\bm{v}_l- \bm{Y}^\top \bm{g}_l\|_2^2 \label{lem2_1}\\
= & - \mathrm{tr}\bm{Y}^\top \Big(\bm{W}
+\sum_{l \in \varepsilon}\bm{g}_l\bm{\lambda}_l^\top
+ \rho \sum_{l \in \varepsilon}\bm{g}_l \bm{v}_l^{\top} \Big)
+ \frac{\rho}{2} \mathrm{tr} \Big( \bm{Y}^\top \sum_{l \in \varepsilon}\bm{g}_l\bm{g}_l^\top \bm{Y} \Big) \label{lem2_2}\\
\leq& - \mathrm{tr}\bm{Y}^\top \Big(\bm{W}
+\sum_{l \in \varepsilon}\bm{g}_l\bm{\lambda}_l^\top
+ \rho \sum_{l \in \varepsilon}\bm{g}_l \bm{v}_l^{\top} \Big)
+2 \omega - 2\mathrm{tr}(\bm{Y}^\top (\omega\bm{I}-\bm{C})\bm{Q}) - \mathrm{tr} \big( \bm{Q}^\top \bm{C}\bm{Q} \big) \label{lem2_3}\\
= & - \mathrm{tr}\bm{Y}^\top \Big( \bm{W}
+\sum_{l \in \varepsilon}\bm{g}_l\bm{\lambda}_l^\top
+ \rho \sum_{l \in \varepsilon}\bm{g}_l\bm{v}_l^{\top} \Big)
- 2\mathrm{tr}(\bm{Y}^\top(\omega\bm{I}-\bm{C})\bm{Q}) + \mathrm{const} \nonumber\\
= & - \mathrm{tr}\bm{Y}^\top \Big(\bm{W}
+\sum_{l \in \varepsilon}\bm{g}_l\bm{\lambda}_l^\top
+ \rho \sum_{l \in \varepsilon}\bm{g}_l \bm{v}_l^{\top} 
+ 2(\omega\bm{I}-\bm{C})\bm{Q}\Big) + \mathrm{const}.  \label{lem2_4}
\end{align}

From Eq. (\ref{lem2_1}) to Eq. (\ref{lem2_2}), the transformation from Eq. (\ref{okikae_y}) to Eq. (\ref{tenkai_y}) is used. 
Next, from Eq. (\ref{lem2_2}) to Eq. (\ref{lem2_3}), the right-hand side of Eq. (\ref{maj_ymat}), derived using the majorizing function, was applied to the second term of Eq. (\ref{lem2_2}). 
Then, Eq. (\ref{lem2_4}) can be explained as a linear term of $\bm{Y}$.

\end{proof}

The left-hand side of Eq. (\ref{lemma2}) are terms related to $\bm{Y}$ from Eq. (\ref{prop_lag}). 

\end{lem}

\begin{dfn}
Eq. (\ref{lem2_4}) is defined as the majorizing function for the objective function:

\begin{align}
M(\bm{Y}|\bm{Q}) = -\mathrm{tr}\bm{Y}^\top \Big(\bm{W}
+\sum_{l \in \varepsilon}\bm{g}_l\bm{\lambda}_l^\top
+ \rho \sum_{l \in \varepsilon}\bm{g}_l \bm{v}_l^{\top} 
+ 2(\omega\bm{I}-\bm{C})\bm{Q}\Big) + \mathrm{const}.
\label{maj_prop}
\end{align}

\end{dfn}

When $\bm{Q} = \bm{Y}$, Eq. (\ref{maj_prop}) is equal to the function in Eq. (\ref{prop_lag}). 
${\bf Proposition \ \ref{prop2}}$ states that the problem of minimizing Eq. (\ref{prop_lag}) can be expressed in Eq. (\ref{proc}).

\begin{prop}\label{prop2}

Given $\bm{W}, \bm{g}_l, \bm{\lambda}_l, \bm{v}_l, \bm{C}, \bm{Q},$ and $\omega$, the minimization problem of the majorizing function $M(\bm{Y}|\bm{Q})$ is equivalent to solving the orthogonal Procrustes problem for Eq. (\ref{proc}).

\begin{align*}
\|\bm{Y}-\bm{D} \|^2_F \rightarrow \min
\end{align*}

s.t. $\bm{Y}^\top \bm{Y}=\bm{I}$,\\

where

\begin{align*}
\bm{D} = \frac{1}{2}\bigg( \bm{W} +  \sum_{l \in \varepsilon}\bm{g}_l\bm{\lambda}_l^\top
+ \rho \sum_{l \in \varepsilon}\bm{g}_l \bm{v}_l^{\top}
+  2(\omega\bm{I}-\bm{C})\bm{Q}\bigg).
\end{align*}

\begin{proof}

From Eq. (\ref{procru}) and the constraint $\bm{Y}^\top \bm{Y}=\bm{I}$,

\begin{align}
\|\bm{Y}-\bm{D} \|^2_F = \mathrm{tr}(\bm{Y}^\top\bm{Y}) - 2\mathrm{tr}(\bm{Y}^\top \bm{D}) + \mathrm{const} = (k-1) - 2\mathrm{tr}(\bm{Y}^\top \bm{D}) + \mathrm{const}.
\label{prf_prop2}
\end{align}

Since the first term is a constant value due to the constraints, the minimization problem in Eq. (\ref{prf_prop2}) is the same as the maximization problem of $2\mathrm{tr}(\bm{Y}^\top \bm{D})$ in Eq. (\ref{prf_prop2}). 
Therefore, the minimization problem of Eq. (\ref{prf_prop2}) is equivalent to the maximization problem of $M(\bm{Y}|\bm{Q})$.

\end{proof}
\end{prop}

\subsection{Update $\vect{Y}^\dagger$ \label{subsec:algY}}

In this subsection, the updated formula for $\bm{Y}$, $\bm{V}$, and $\bm{\Lambda}$ in $\bm{Y}^\dagger$ by using the ADMM algorithm are described.

\noindent
{\bf Update $\bm{Y}$}

The updated formula of $\bm{Y}$ in ADMM algorithm is shown in \ref{subsec:maj_socdc} as 

\begin{align*}
\bm{Y} \leftarrow \bm{L}\bm{R}^\top,
\end{align*}

\noindent
where $\bm{L}$ and $\bm{R}$ come from the singular value decomposition from $\bm{D}$ as the following:

\begin{align*}
\bm{D} = \frac{1}{2}\bigg(\bm{W} +  \sum_{l \in \varepsilon}\bm{g}_l\bm{\lambda}_l^\top
+ \rho \sum_{l \in \varepsilon}\bm{g}_l \bm{v}_l^{\top}
+  2(\omega\bm{I}-\bm{C})\bm{Q}\bigg)= \bm{L}\bm{\Sigma}\bm{R}^\top
\end{align*}

from {\bf Proposition \ \ref{prop2}}.
\noindent
{\bf Update $\bm{V}^\dagger$}


\begin{prop}\label{prop3}

Given $\bm{s}_l, \psi_l \ (\psi_l >0)$, the updated formula of $\bm{v}_l$ is 

\begin{align}
\bm{v}_l \leftarrow
\begin{cases}
\bm{\bm{s}_l} \Big(1-  \frac{\psi_l}{\|\bm{s}_l \|_2} \Big), &\ (\|\bm{\bm{s}_l} \|_2 > \psi_l) \\
\bm{0}, &\ (\| \bm{\bm{s}_l}\|_2 \leq \psi_l) 
\end{cases}
\label{update_v}
\end{align}

\noindent
where $\bm{s}_l = \bm{v}_l -(\psi_l(\bm{v}_l-\bm{q}_l))$, $\bm{q}_l = \bm{y}_{i} - \bm{y}_{j} -\rho^{-1} \bm{\lambda}_l$, and $\psi_l=\frac{\gamma \alpha_l}{\rho}$.

\begin{proof}

The updated formula of $\bm{V}$ can be solved in the same manner as \cite{chiadmm}. 
It is derived using the proximal gradient method. 
First, for a given $l$ in the terms related to $\bm{V}$ in Eq. (\ref{prop_lag}),

\begin{align*}
& \bm{\lambda}_l^\top(\bm{v}_l- \bm{y}_{i}+ \bm{y}_{j}) 
 +\frac{\rho}{2}\|\bm{v}_l - \bm{y}_{i} + \bm{y}_{j}\|_2^2 
 + \gamma \alpha_l \|\bm{v}_l\|_2
\end{align*}

First, multiply $\rho^{-1}$ to Eq. (\ref{prop_lag}).

\begin{align*}
& \rho^{-1}\bm{\lambda}_l^\top(\bm{v}_l- \bm{y}_{i}+ \bm{y}_{j}) 
 +\frac{1}{2}\|\bm{v}_l - \bm{y}_{i} + \bm{y}_{j}\|_2^2 
 + \frac{\gamma \alpha_l}{\rho}\|\bm{v}_l\|_2 \\
=& \frac{1}{2}\big( (\bm{v}_l - \bm{y}_i + \bm{y}_j)^\top (\bm{v}_l - \bm{y}_i + \bm{y}_j) \big) 
+ \rho^{-1}\bm{\lambda}_l^\top(\bm{v}_l - \bm{y}_i + \bm{y}_j) 
+ \frac{\gamma \alpha_l}{\rho}\|\bm{v}_l\|_2\\
=& \frac{1}{2} \bigg( \big((\bm{v}_l - \bm{y}_i + \bm{y}_j) + \rho^{-1}\bm{\lambda}_l\big)^\top \big((\bm{v}_l - \bm{y}_i + \bm{y}_j) + \rho^{-1}\bm{\lambda}_l\big)\bigg)
-\frac{1}{2}\|\rho^{-1}\bm{\lambda}_l \|_2^2
+ \frac{\gamma \alpha_l}{\rho}\|\bm{v}_l\|_2
\end{align*}

As for the terms related to $\bm{v}_l$:

\begin{align*}
& \frac{1}{2} \bigg( \big((\bm{v}_l - (\bm{y}_i - \bm{y}_j - \rho^{-1}\bm{\lambda}_l\big)^\top \big((\bm{v}_l - (\bm{y}_i - \bm{y}_j - \rho^{-1}\bm{\lambda}_l)\big)\bigg)
+ \frac{\gamma \alpha_l}{\rho}\|\bm{v}_l\|_2\\
=& \frac{1}{2}\| \bm{v}_l - (\bm{y}_i - \bm{y}_j -\rho^{-1} \bm{\lambda}_l) \|_2^2 
+ \frac{\gamma \alpha_l}{\rho}\|\bm{v}_l\|_2.
\end{align*}

Therefore, the updated formula for $\bm{v}^\dagger_l$ can be obtained by solving the following minimization problem.

\begin{align}
\argmin_{\bm{v}_l}\frac{1}{2}\| \bm{v}_l - (\bm{y}_{i} - \bm{y}_{j} - \rho^{-1}\bm{\lambda}_l)\|_2^2 
+ \frac{\gamma \alpha_l}{\rho}
\|\bm{v}_l\|_2.
\label{v_upd_ob}
\end{align}

Then, Eq. (\ref{v_upd_ob}) is divided into the differentiable terms and the other terms, similar to the case of the updated formula of $\bm{B}$ in Eq. (\ref{prox_pro}).

\begin{align*}
\begin{cases}
\mathscr{F}(\bm{v}_l) = \frac{1}{2}\| \bm{v}_l - (\bm{y}_{i} - \bm{y}_{j} - \rho^{-1}\bm{\lambda}_l)\|_2^2 \\
\mathscr{G}(\bm{v}_l) = 
\frac{\gamma \alpha_l}{\rho} \|\bm{v}_l\|_2 
\end{cases}
\end{align*}

The gradient for $\mathscr{F}(\bm{v}_l)$ is calculated as follows and set as $\bm{s}_l$: 

\begin{align*}
\bm{s}_l \equiv \bm{v}^\dagger_l- \psi_l (\bm{v}^\dagger_l - \bm{q}_l).
\end{align*}

From this, the proximal operator of $\mathscr{G}(\bm{v}_l)$ can be derived as

\begin{align}
{\bf prox}_{\psi_l \mathscr{G}}(\bm{s}_l)= \argmin \Big( \psi_l \| \bm{v}_l \|_2  + \frac{1}{2}\| \bm{v}_l -\bm{s}_l \|_2^2\Big).
\label{prox_v}
\end{align}

Similar to Eq. (\ref{prox_w}), when $\psi_l \neq 0$, Eq. (\ref{prox_v}) is obtained by differentiating with respect to $\bm{v}_l $, and setting the result equal to $\bm{0}$.

\begin{align*}
\bm{v}_l  - \bm{s}_l + \psi_l  \frac{\bm{v}_l }{\| \bm{v}_l \|_2}  =& \bm{0}\\
\Longleftrightarrow \hspace{40mm}
\bm{v}_l  \Bigg( 1 + \psi_l \frac{1}{\|\bm{v}_l  \|_2}\Bigg) =& \bm{s}_l \tag{iii}\\
\Longleftrightarrow \hspace{9mm}
\Bigg( 1 + \psi_l \frac{1}{\|\bm{v}_l  \|_2}\Bigg)^\top \bm{v}_l ^{\top} \bm{v}_l  \Bigg( 1 + \psi_l \frac{1}{\|\bm{v}_l  \|_2}\Bigg)=& \bm{s}_l^\top \bm{s}_l\\
\Longleftrightarrow \hspace{34.3mm}
\Bigg( 1 + \psi_l \frac{1}{\bm{v}_l  \|_2}\Bigg)^2 \|\bm{v}_l \|^2_2 =& \|\bm{s}_l \|^2_2\\
\Longleftrightarrow \hspace{35mm}
\Bigg( 1 + \psi_l \frac{1}{\|\bm{v}_l  \|_2}\Bigg) \|\bm{v}_l  \|_2 =& \|\bm{s}_l \|_2\\
\Longleftrightarrow \hspace{51mm}
\|\bm{v}_l  \|_2 + \psi_l =& \|\bm{s}_l \|_2\\
\Longleftrightarrow \hspace{58.5mm}
\|\bm{v}_l  \|_2 =&  \|\bm{s}_l \|_2 - \psi_l \tag{iv}
\end{align*}

Substitute Eq. (iv) into Eq. (iii).

\begin{align*}
\bm{v}_l \Bigg( 1 + \psi_l \frac{1}{\|\bm{s}_l \|_2 - \psi_l}\Bigg) =& \bm{s}_l \\
\Longleftrightarrow \hspace{5mm}
\bm{v}_l \Bigg( \frac{\|\bm{s}_l \|_2 - \psi_l + \psi_l}{\|\bm{s}_l \|_2 - \psi_l}\Bigg) =& \bm{s}_l \\
\Longleftrightarrow \hspace{34.5mm}
\bm{v}_l =& \bm{\bm{s}_l} \Bigg(1-  \frac{\psi_l}{\|\bm{s}_l \|_2} \Bigg)
\end{align*}

\end{proof}
\end{prop}

\noindent
{\bf Update $\bm{\Lambda}$}

The updated formula for $\bm{\Lambda}$ is based on ADMM, and it is updated for each $l$, which is defined as follows:

\begin{align*}
\bm{\lambda}_l \leftarrow \bm{\lambda}_l + \rho(\bm{v}_l- \bm{y}_{i} + \bm{y}_{j}).
\end{align*}

The detail of the algorithm is shown in ${\rm Algorithm ~\ref{alg1}}$.

{\footnotesize
\begin{spacing}{0.1}
\begin{algorithm}[H]
    \caption{Regularized sparse optimal discriminant clustering}
    \small
\footnotesize
    \label{alg1}
    \begin{algorithmic}[1]    
    \REQUIRE ${\bm X}, {\bm H}_n, k ,\; \eta_1 > 0, \gamma > 0, \rho>0, \alpha_l \ (l \in \varepsilon) \geq 0,$ threshold for this algorithm $\epsilon > 0$; value of RSODC $ L^{(t)}$, value of Eq. (\ref{prop_lag}) $L_{Y}^{(t)}$
    \ENSURE ${\bm \beta}_j \ (j=1, 2, \cdots, p), {\bm Y}, {\bm V}, {\bm \Lambda}$
    \STATE Set $t \leftarrow 1$
    \FOR {$j=1$ to $p$}
    \STATE Set initial values ${\bm \beta_j^{(t)}}$
    \ENDFOR 
    \STATE Set initial ${\bm Y}^{(t)}$, ${\bm V}^{(t)}$ and ${\bm \Lambda}^{(t)}$
    \WHILE{ $L^{(t)} - L^{(t+1)}$ $\geq \epsilon$ }
    \STATE {\bf Update $\bm{B}$}:
    \STATE Update $\bm{\beta}_j^{(t+1)}$ based on Eq. (\ref{upd_beta})
    \STATE {\bf Update $\bm{Y}^\dagger$}:
    \WHILE{ $L_Y^{(t)} - L_Y^{(t+1)}$ $\geq \epsilon$ }
    \STATE {$\bm{Y}^{(t+1)} \leftarrow \bm{L}\bm{R}^\top$ by ${\bf Proposition 2}$}, 
    \STATE Update $\bm{v}_l^{ (t+1)}$ based on Eq. (\ref{update_v}) 
    \STATE $\bm{\lambda}_l^{(t+1)} \leftarrow \bm{\lambda}_l^{(t)} + \rho(\bm{v}_l- \bm{y}_{i} + \bm{y}_{j})$
    \ENDWHILE
    \STATE $\bm{Y}^{\dagger (t+1)} \leftarrow \bm{Y}^{\dagger (t)}$
    \STATE $t \leftarrow (t+1)$
    \ENDWHILE
    \end{algorithmic}
\end{algorithm}
\label{alg}
\end{spacing}
}

\subsection{Cross-validation}
\label{subsec:cv}

RSODC selects $\eta_1, \gamma$ and $\rho$ using the cross-validation method proposed in SODC. This procedure is based on clustering stability \citep{SODC, Sun2013}. 
This cross-validation evaluates the agreement ratio between these two subsets by using Cohen's kappa coefficient \citep{cohen1960coefficient}.  
When the training data samples are randomly split into two halves, denoted as $\bm{X}^{(1)}$ and $\bm{X}^{(2)}$, RSODC is applied to each split with the same candidate values of $\eta_1, \gamma$ and $\rho$. 
Then, we receive two independent $\bm{\hat{B}}$, and name them $\bm{\hat{B}}^{(1)}$ and $\bm{\hat{B}}^{(2)}$. 
Since $\bm{\hat{B}}$ is estimated using the group lasso, elements in the same row are set to zero if the corresponding variable is not important for classification. From $\bm{\hat{B}}^{(1)}$ and $\bm{\hat{B}}^{(2)}$, we create each indicator vector $\bm{\mathscr{I}}_{\bm{\hat{B}}^{(1)}} \in \{0,1 \}^p$ and $\bm{\mathscr{I}}_{\bm{\hat{B}}{(2)}} \in \{0,1 \}^p$ respectively. These vectors assign $1$ to rows that are non-zero and $0$ to rows that are zero. 
A simple example is given below.

\begin{align*}
\widehat{\mathbf{B}}^{(1)} =
\begin{bmatrix}
0 & 0 \\
1.2 & -0.5 \\
0 & 0 \\
-0.3 & 0.7 \\
0 & 0
\end{bmatrix}, \quad
\widehat{\mathbf{B}}^{(2)} =
\begin{bmatrix}
0 & 0 \\
1.6 & 0.9 \\
0.4 & -0.2 \\
-0.3 & 0.7 \\
0 & 0
\end{bmatrix}.
\end{align*}

In this case, the indicator vector for each estimated parameter is constructed as

\begin{align*}
\bm{\mathscr{I}}_{\bm{\hat{B}}^{(1)}} =
\begin{bmatrix}
0  \\
1 \\
0 \\
1 \\
0
\end{bmatrix}, \quad
\bm{\mathscr{I}}_{\bm{\hat{B}}^{(2)}} =
\begin{bmatrix}
0  \\
1  \\
1 \\
1 \\
0
\end{bmatrix}.
\end{align*}

Using these two indicator vectors, the Kappa coefficients are calculated. Follow the rule of the conventional SODC, when $\bm{\mathscr{I}}_{\bm{\hat{B}}^{(1)}}$ and $\bm{\mathscr{I}}_{\bm{\hat{B}}^{(2)}}$ are either both non-selected or both full selected, the Kappa coefficients are treated as $-1$. 
This procedure is repeated multiple times for each pattern candidate of $\eta_1, \gamma$ and $\rho$. In this study, the number of repetitions was set to $10$.
The combination of parameters yielding the highest mean kappa coefficients is selected.

\section{Numerical Simulation}
\label{sec:sim}

\subsection{Simulation design}
\label{subsec:simdesign}

We implemented five numerical simulations to assess the performance of RSODC. In Simulation 1, the estimation of RSODC is compared with other methods, and the influence of the settings in RSODC is examined in Simulation 2, 3, 4, and 5. 
First, we explain the design of numerical simulations. 

\subsubsection{Simulation 1}
\label{subsubsec:simdesign1}

We first compare the performance of RSODC with the compared methods. 
The settings are based on a modified version of \cite{SODC}.
The data matrix $\bm{X} \in \mathbb{R}^{n \times p}$ is generated to contain the true clustering structure. 
The number of subjects $n$ is $60, 96$ and $156$, and the number of covariates $p$ is $20$, $50$, $80$ and $100$. The number of clusters $k$ is $3$ and $4$. 
The evaluation index is Adjusted Rand Index (ARI) \citep{hubert1985comparing} between the estimated clustering structure and the true clustering structure.

Covariates corresponding to cluster $\ell$ are generated from $\bm{X}_\ell \sim N(\bm{\mu}_\ell, \tilde{\bm{\Sigma}}) \ (\ell=1, 2, \cdots, k)$, where $\bm{\mu}_\ell \in \mathbb{R}^p$ is the mean vector of cluster $\ell$, and 

\begin{align*}
\tilde{\bm \Sigma}
= \left(
\begin{array}{lll}
\bm{\Sigma}^{\S} & {\bm O_{q,c^*}} & {\bm O_{q,(p-q-c^*)}}\\
{\bm O_{c^*,q}} & \bm{\Sigma}^\ddag & \bm O_{c^*,(p-q-c^*)}\\
{\bm O_{(p-q-c^*),q}} & {\bm O_{(p-q-c^*),c^*}} & \bm{I}_{(p-q-c^*)}
\end{array}
\right).
\end{align*}

$\tilde{\bm \Sigma}$ is the covariance matrix and $\bm{I}_{(p-q-c^*)}$ is the identity matrix. 
$\bm{X}$ contains the informative variable from the first to $q$th, where $q$ is the number of informative covariates, and $q$ is set as $2$. 
First, $\Sigma^{\S}$ is the covariance within the informative variables, which includes the true clustering structure. It is set as 

\begin{align}
\Sigma^{\S} = (1-\xi)\bm{I}_q +\xi\bm{1}_q\bm{1}_q^\top \in \mathbb{R}^{q \times q},
\label{sigma_info}
\end{align}

where $\xi$ is the variance of the informative covariates, whose settings are denoted in Factor 5. 
$\bm{I}_q$ is identity matrix and $\bm{1}_q = (1, 1, \cdots, 1)^\top$ is a vector of $1$. 

Next, for the remaining of the variables, the non-informative covariates $(p-q)$, we set two different parts: non-informative covariates with higher correlation and those with no correlation.
For non-informative covariates with higher correlations, we set:

\begin{align*}
\Sigma^\ddag = (1-\xi^\dagger)\bm{I}_{c^*} 
+ \xi^\dagger \bm{1}_{c^*}\bm{1}^\top_{c^*} \in \mathbb{R}^{c^* \times c^*}.
\end{align*}

Here, $\xi^\dagger = 0.6$.
This corresponds to the $(q+1)$ th to $(q+c^*)$th variables in $\bm{X}$. 
$c^*$ is the number of non-informative covariates with higher correlation. $c^*$ is set to $12$ for $p=20$, $24$ for $p=50$, $36$ for $p=80$ and $48$ for $p=100$. 
$\bm{I}_{c^*}$ is the identity matrix and $\bm{1}_{c^*}$ is a vector of $1$.
The rest, from $(q+c^*+1)$th to $p$th non-informative covariates, are generated from the standard normal distribution. 

Next, the mean vectors of each cluster $\bm{\mu}_\ell = (\bm{m}_\ell^\top, \bm{0}^\top)$ are determined as follows:

\begin{align}
\bm{m}_1 =&  \vartheta(-\bm{1}^\top_{q/2},\bm{1}^\top_{q/2})^\top, \\
\bm{m}_2 =&\vartheta\bm{1}_q,\\
\bm{m}_3 =&\vartheta(\bm{1}^\top_{q/2},-\bm{1}^\top_{q/2})^\top, \ {\rm and} \\
\bm{m}_4 =& \vartheta(-\bm{1}_q), 
\label{mean_clus}
\end{align}

where $\vartheta$ is a parameter that sets the location of the cluster mean, which will be indicated in Factor 4.

The number of total patterns is 
$5$ (Factor 1) $\times$ $3$ (Factor 2)  $\times 4$ (Factor 3) $\times 2$ (Factor 4) $\times 3$ (Factor 5) $\times 2$ (Factor 6) $=720$. 
We randomly generate training data and test data, and repeat the calculation $100$ times for each pattern. 
For evaluation, Adjusted Rand Index (ARI) between the estimated clustering structure and the true clustering structure is used to assess how well they match the true clusters. 
For the proposed method and SODC, the ARI was calculated from $\bm{H}_n \bm{X}\hat{\bm{B}}$, where $\hat{\bm{B}}$ denotes the estimated $\bm{B}$. 
To calculate ARI, {\choosefont{pcr}mclust} package in R software \citep{r_mclust} is used.

\noindent
{\bf Factor 1: Methods}

We compare four methods, and the four compared methods are applied: sparse optimal discriminant clustering (SODC), tandem clustering \citep{Arbietandem} in $(k-1)$ dimensions, reduced $k$-means \citep{de1994k} in $(k-1)$ dimensions, and Factorial $k$-means \citep{VICHI200149} in $(k-1)$ dimensions. We use Eq. (\ref{sodc}) for SODC. In this setting, we set $\eta_2 =0$ in both RSODC and SODC.  

For reduced $k$-means, we use {\choosefont{pcr}clustrd} package in R software \citep{rdm_package}. 
The tuning parameters $\eta_1$, $\gamma$, and $\rho$ in the proposed method are selected by modified cross-validation based on the idea of clustering stability explained in \ref{subsec:cv}.
For the weight of the penalty term for $\bm{Y}$, we adapted $\tau=0.1$ and $\delta=25$. The parameter $\nu$ in the updated formula of $\bm{B}$ is set as $0.001$. 
For SODC, $\eta_1$ is selected in the same manner. 

\noindent
{\bf Factor 2: Number of subjects}

The number of subjects is set as $n=60, 96$, and $156$.

\noindent
{\bf Factor 3: Covariate variable}

The number of covariate variable is set as $p=20$, $50$, $80$ and $100$.

\noindent
{\bf Factor 4: Number of cluster}

The number of cluster is $k=3$ and $4$.

\noindent
{\bf Factor 5: Parameter $\vartheta$ for informative covariate variable}

$\vartheta$ is distance between cluster centroids. $\vartheta$ is set as $1.4, 2.0$, and $2.2$.

\noindent
{\bf Factor 6: Parameter $\xi$ for informative covariate variable}

$\xi$ is variance of informative covariate variables, and set as $\xi=0$ and $0.5$.\\

In RSODC, the initialization of $\bm{B}$ is generated from $N(0,1)$, the initial $\bm{Y}$ is set as $(\bm{X}\bm{X}^\top)^{-1}$, $\bm{v}_l = \bm{y}_i-\bm{y}_j$ from initial $\bm{Y}$ and $\bm{\Lambda}$ is set as $\bm{O}^{((n(n-1))/2)\times(k-1)}$. The initial parameters in SODC are generated similarly.

For the candidates of parameters $\eta_1$, $\gamma$ and $\rho$, we have $\eta_1 = (0.1, 0.5, 1, 1.5, 2, 2.5, 3), \gamma = (0.001, 0.003, 0.005, 0.007, 0.01)$ and $\rho = (0.01, 0.03, 0.05 ,0.07, 0.1)$ respectively,  and select the combination of $\gamma$ and $\rho$ as $\gamma/\rho < 1$ to keep the update of $\bm{V}$ stable. 
In SODC, the candidates $\eta_1 = (0.1, 0.3, 0.5, 0.7, 1, 1.5, 2, 2.5, 3, 3.5, 4)$. 
The number of cross-validation is $10$ in both RSODC and SODC due to calculation cost.

\subsubsection{Simulation 2}
\label{subsubsec:simdesign2}

In this simulation, we examine the effect on the estimation of RSODC by selecting each tuning parameter $\eta_1, \gamma$ and $\rho$. 
RSODC is calculated $100$ times with each pattern of the parameters to compare ARI. 
For the data matrix $\bm{X}$, the number of subjects is $n=60$, the number of covariate variables is $p=20$, and the number of clusters is $k = 3$. The generation of $\bm{X}_\ell$ is the same as in Simulation 1. 
In data generation, the parameter of the covariance for the informative variables, $\xi$ in Eq. (\ref{sigma_info}) is set as $0.5$, and the parameter for the mean vector of each cluster, $\theta$, in Eq. (\ref{mean_clus}) is set to $2.2$. The candidates for the tuning parameters $\eta_1, \gamma$, and $\rho$ are the same as those in Simulation 1. The other parameter settings are also the same as in Simulation 1.

\subsubsection{Simulation 3}
\label{subsubsec:simdesign3}

This simulation evaluates the estimated results based on the selected $k$ to assess whether they match the true number of clusters.
We calculate gap statistics \citep{gapstat} using $\bm{B}$ estimated based on each cluster candidate, and examine how often $\bm{B}$ calculated using the true number of clusters is selected out of $100$ trials. 
The data setting is the same as in Simulation 2 and the true cluster is $2$ and $3$.

We explain the procedures for the evaluation. First, RSODC is calculated with initial parameters such as $\bm{Y}$ and $\bm{V}$ generated based on selected clusters. The candidates for the number of clusters are $k=2$ to $9$. 
Secondly, the gap statistics is applied to $\bm{X}^\dagger (=\bm{H}_n \bm{X} \bm{\hat{B}}$) using {\choosefont{pcr}cluster} package in R software \citep{r_cluster}. In this function, the number of Monte Carlo samples is set to $100$. 
In RSODC, it is necessary to set the dimension to be reduced beforehand for estimation.
Therefore, the calculated gap of each selected cluster and its standard error (SE) of the gap are used to compare the more appropriate number of clusters. 
The appropriate number of clusters is determined by the minimum of gap($k$) calculated based on the following values: gap($k$) $\geq$ gap($k+1$) - SE of gap($k+1$). 
This procedure is repeated $100$ times. 
For each cluster candidate, the cross-validation is performed, and the patterns of the parameters $\eta_1, \gamma$ and $\rho$ are set similarly to those in Simulation 1 and 2.

\subsubsection{Simulation 4}
\label{subsubsec:simdesign4}

This simulation focuses on the influence on the estimation by changing two parameters for weight $\alpha_l$ to the regularization of $\bm{V}$ in Eq. (\ref{alpha_weight}). 
The candidates of these parameters are the following: $\tau = (0.001,0.005, 0.01, 0.05, 0.1)$ and $\delta = (5,10,15,20,25,30, 35,40,45,50,55)$. The parameters $\eta_1, \gamma$, and $\rho$ employ the results of cross-validation in Simulation 1. The data settings are the same as in Simulation 2, and the calculation is repeated $100$ times.

For the estimation index, we evaluate ARI, the median of the calculation time, and the median of the number of convergences. In addition, we also examine the sensitivity and specificity of $\bm{\hat{B}}$. 
The ideal situation is that the informative variables are non-sparse and the non-informative variables are sparse. 
The sensitivity and specificity are defined as follows:\\ 

\begin{align*}
Sensitivity  = \frac{{\rm the \ number \ of \ nonzero \ elements \ corresponding \ to \ informative \ variables }}{2 \times (k-1)},
\end{align*}

\begin{align*}
Specificity  = \frac{{\rm the \ number \ of \ zero \ elements \ corresponding \ to \ non-informative \ variables }}{58 \times (k-1)}.
\end{align*}

If the value of sensitivity is high, it means that informative variables are estimated as non-zero and the required information is correctly captured. A high value of specificity indicates that unnecessary information is estimated as zero and that the information is correctly discarded.

\subsubsection{Simulation 5}
\label{subsubsec:simdesign5}

In this simulation, we examine the effects on the estimation by the initialization of $\bm{B}$. RSODC is performed with the randomly generated $\bm{B}$ that follows a normal standard distribution $N(0,1)$ for $100$ trials. 
The evaluation index includes ARI, the calculation time, and the number of the convergences. 
The data setting follows Simulation 2, and the same tuning parameters selected in Simulation 1 are applied.

\subsection{Simulation results}
\label{subsec:simresult}

\begin{figure}[htbp]
\begin{center}
\includegraphics[scale=0.45]{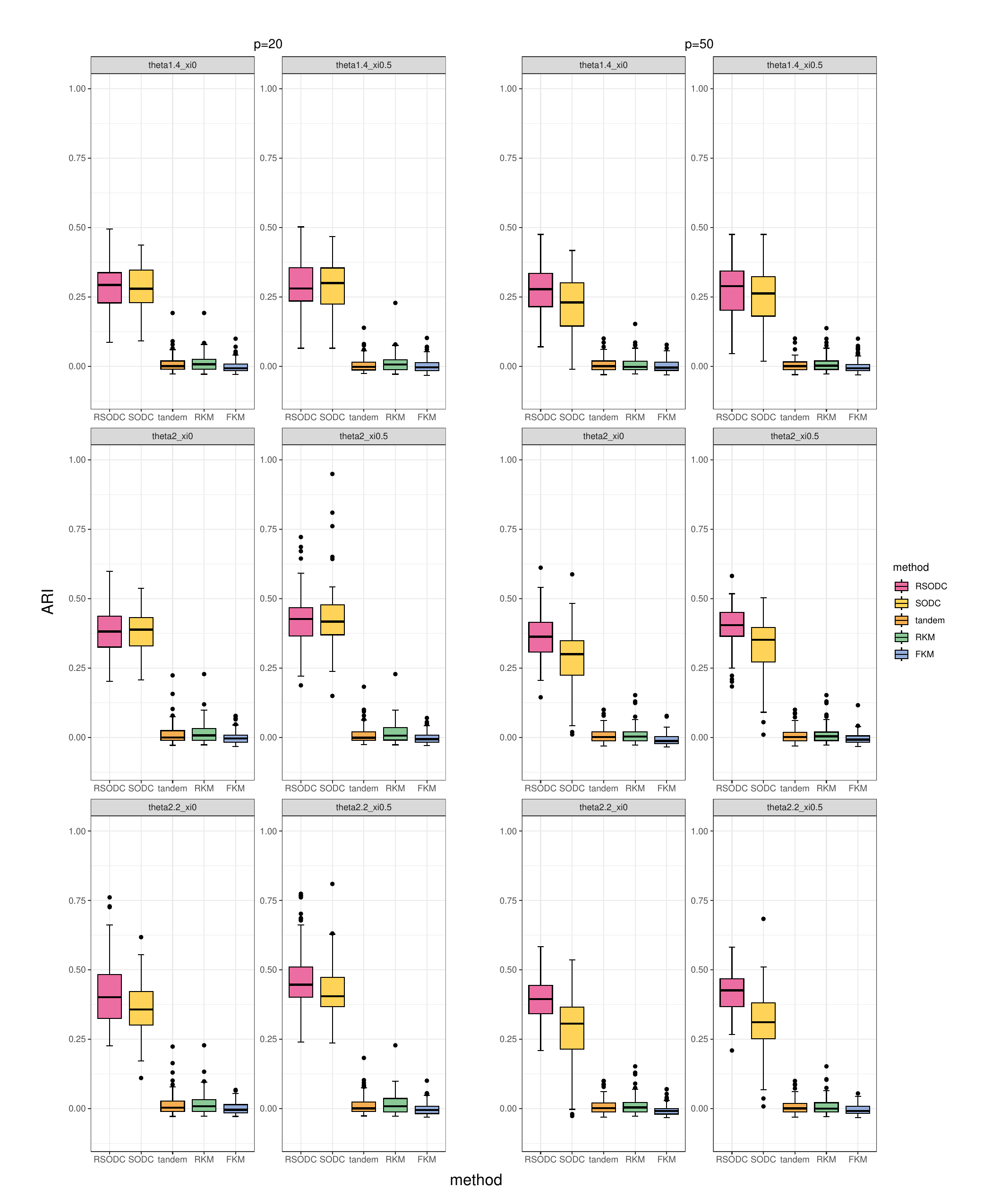}
\caption{Results of ARI in $k=3, n=60, p=20$ and $50$. The vertical axis shows ARI, and the horizontal axis shows the method. In pattern, "theta" presents the the distance between cluster centroids in informative variables, and "xi" represents the variance of informative variables. For the method, "tandem" refers to tandem clustering, "RKM" denotes reduced $k$-means, and "FKM" denotes factorial $k$-means.}
\label{sim_ARIk3n60p2050}
\end{center}
\end{figure}
%
%
\begin{figure}[htbp]
\begin{center}
\includegraphics[scale=0.45]{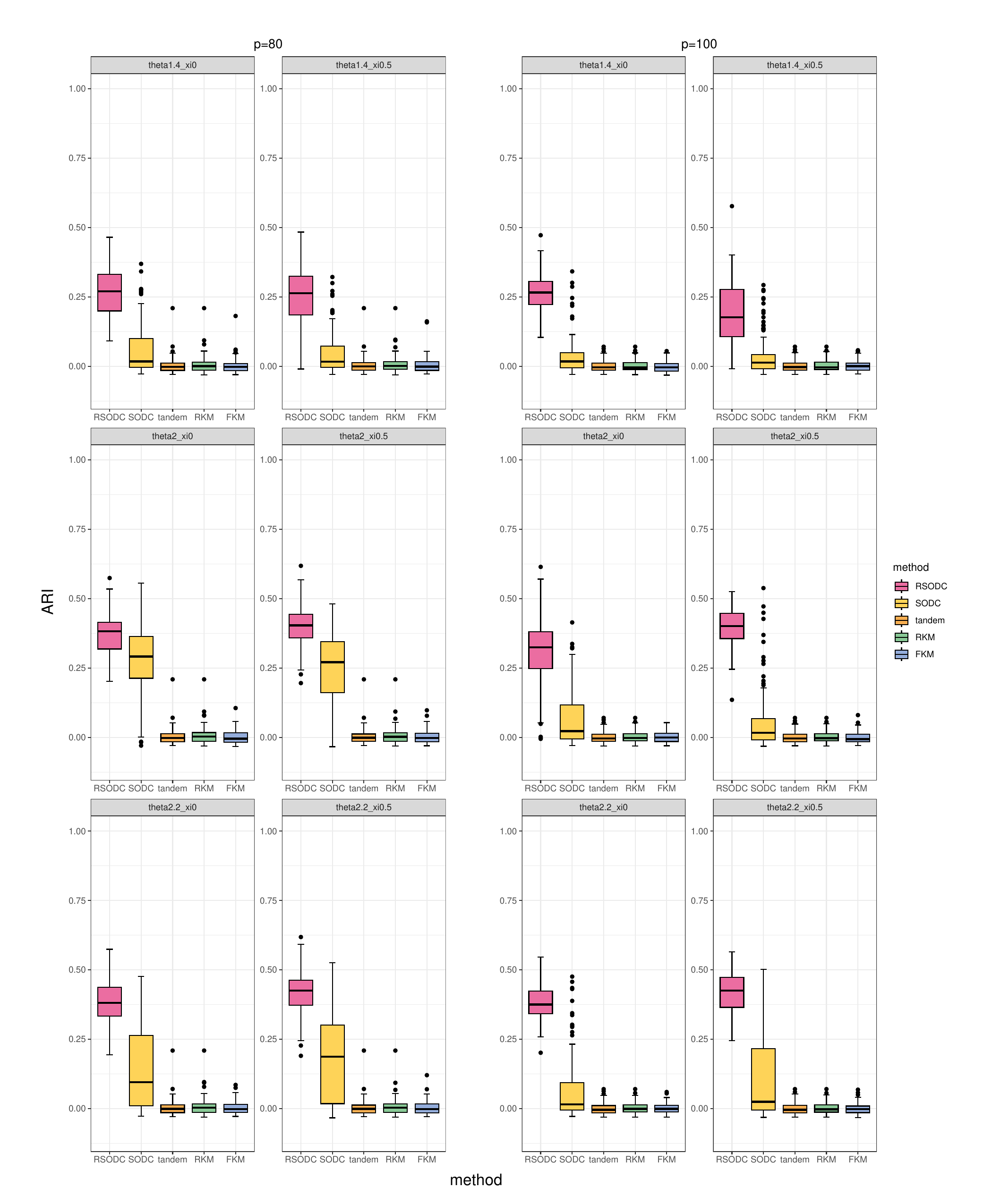}
\caption{Results of ARI in $k=3, n=60, p=80$ and $100$.}
\label{sim_ARIk3n60p80100}
\end{center}
\end{figure}
%
%
\begin{figure}[htbp]
\begin{center}
\includegraphics[scale=0.45]{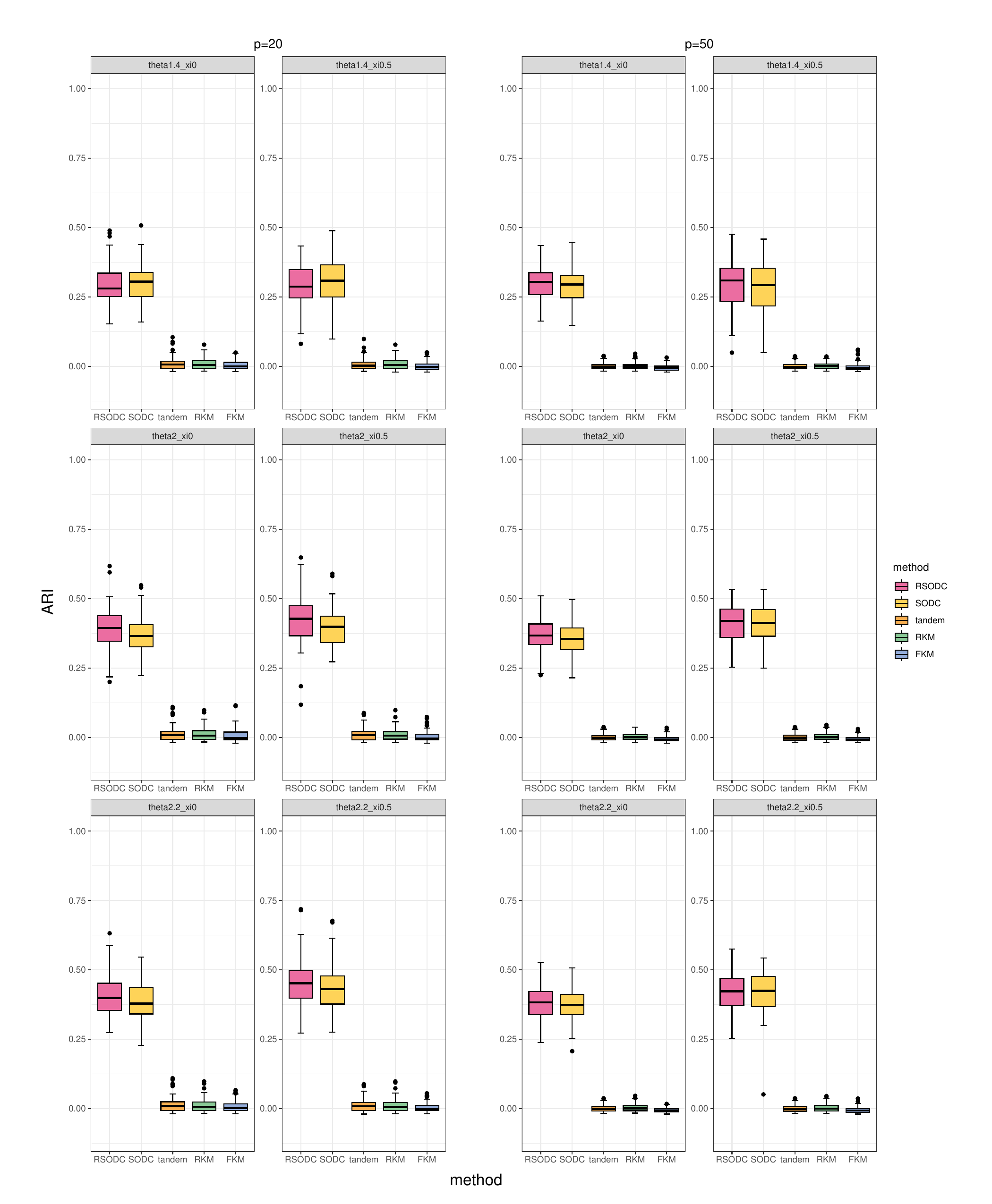}
\caption{Results of ARI in $k=3, n=96, p=20$ and $50$. }
\label{sim_ARIk3n96p2050}
\end{center}
\end{figure}
%
%
\begin{figure}[htbp]
\begin{center}
\includegraphics[scale=0.45]{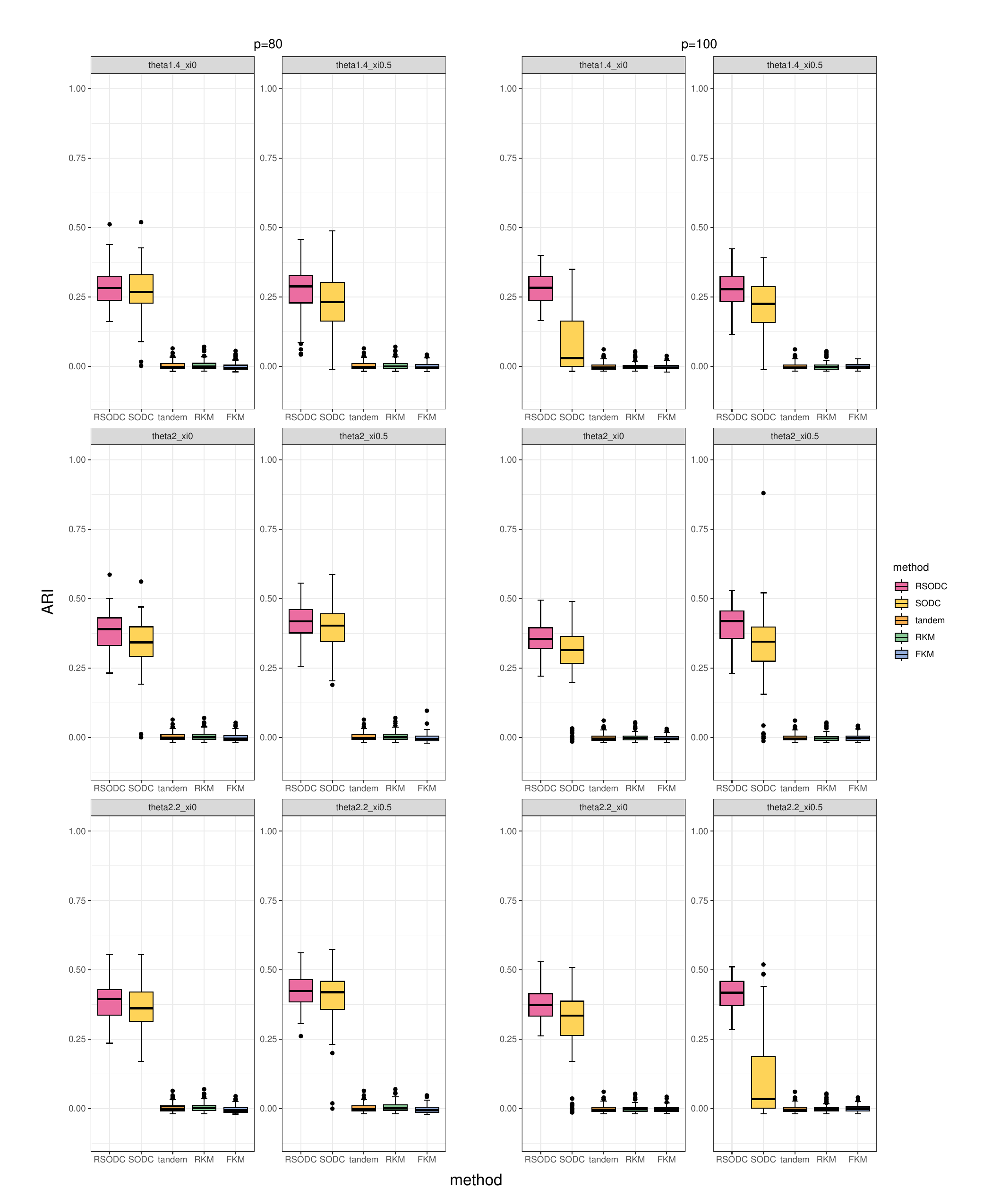}
\caption{Results of ARI in $k=3, n=96, p=80$ and $100$. }
\label{sim_ARIk3n96p80100}
\end{center}
\end{figure}
%

%
\begin{figure}[htbp]
\begin{center}
\includegraphics[scale=0.45]{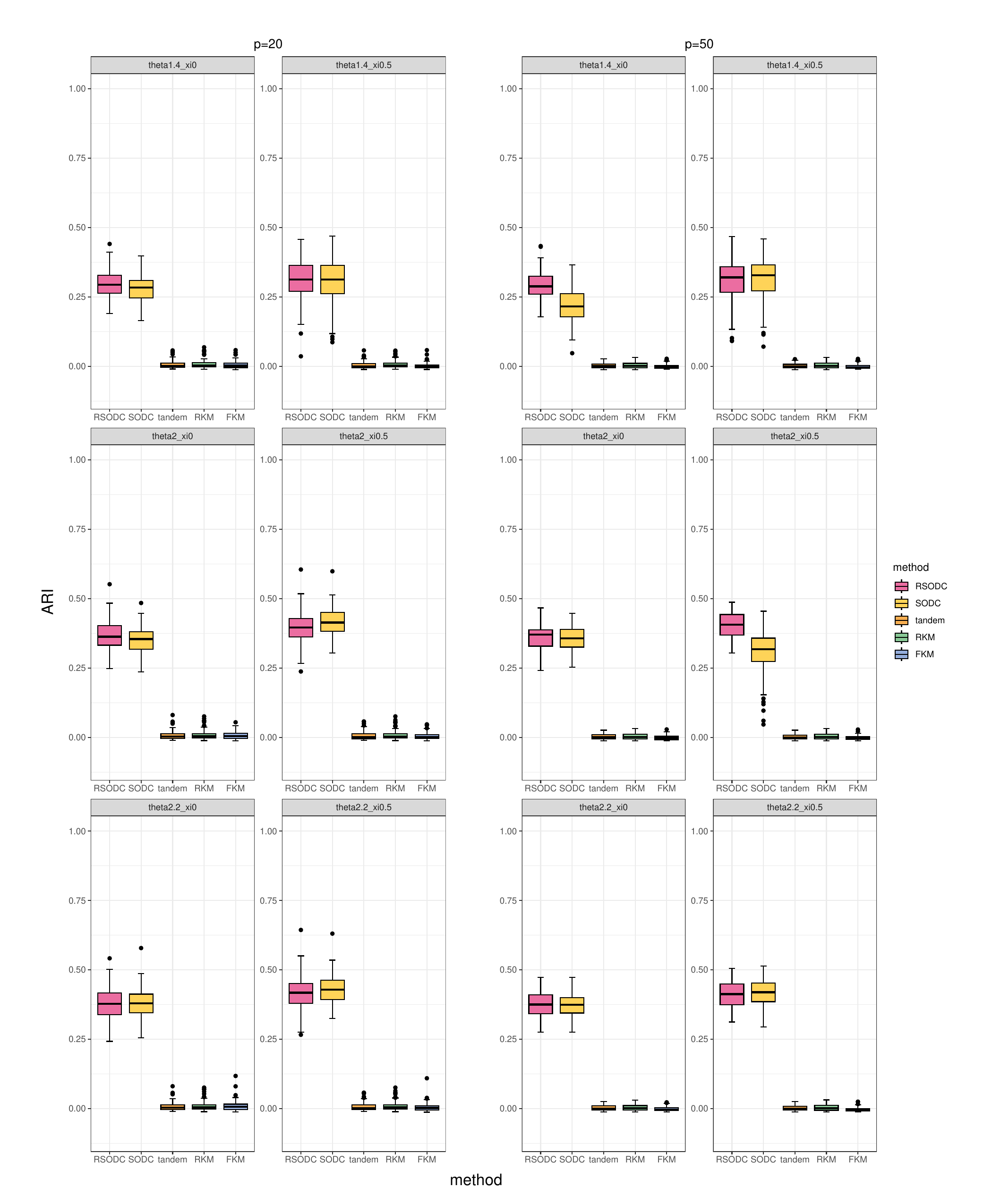}
\caption{Results of ARI in $k=3, n=156, p=20$ and $50$. }
\label{sim_ARIk3n156p2050}
\end{center}
\end{figure}
%
%
\begin{figure}[htbp]
\begin{center}
\includegraphics[scale=0.45]{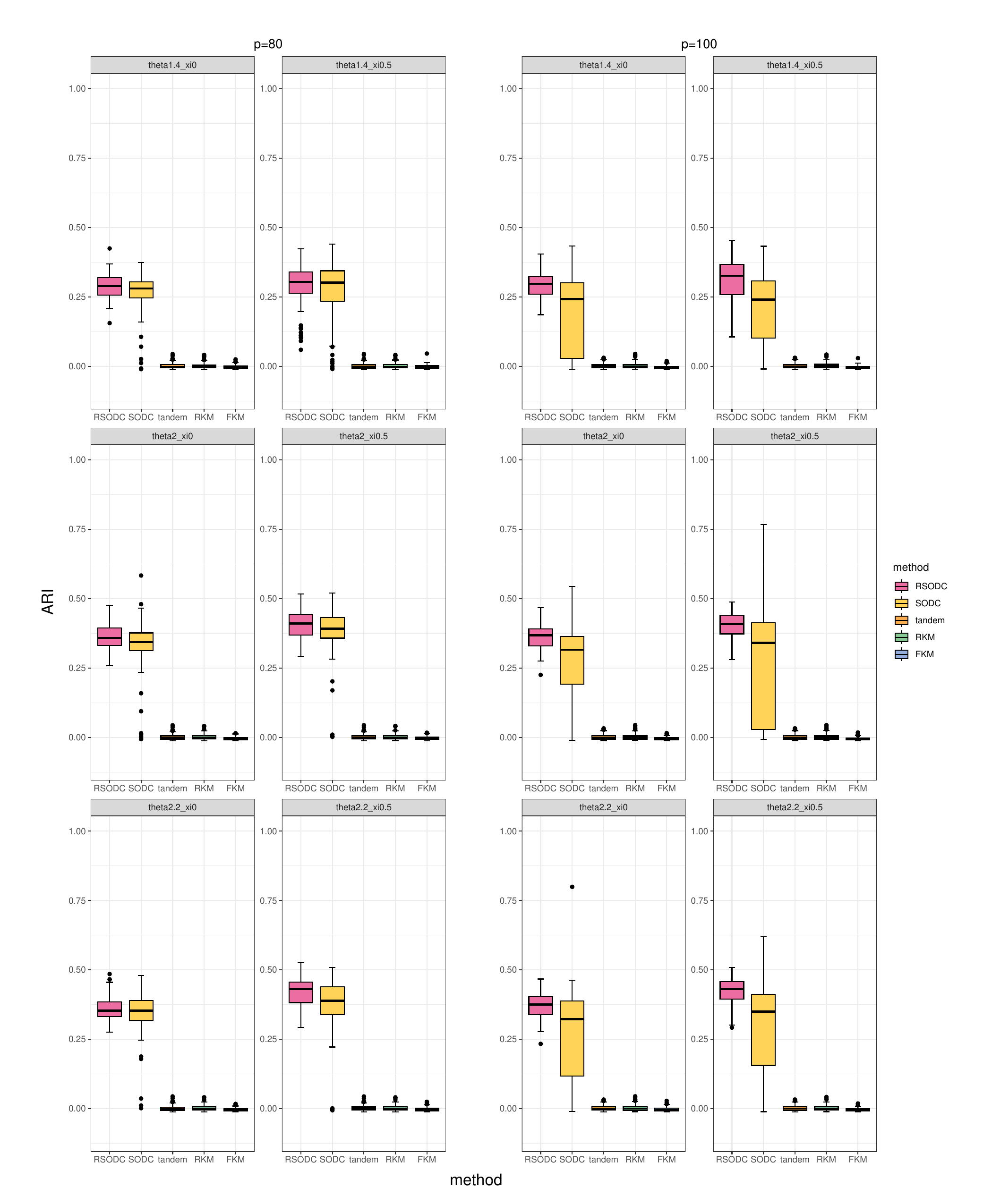}
\caption{Results of ARI in $k=3, n=156, p=80$ and $100$. }
\label{sim_ARIk3n156p80100}
\end{center}
\end{figure}
%

\begin{figure}[htbp]
\begin{center}
\includegraphics[scale=0.45]{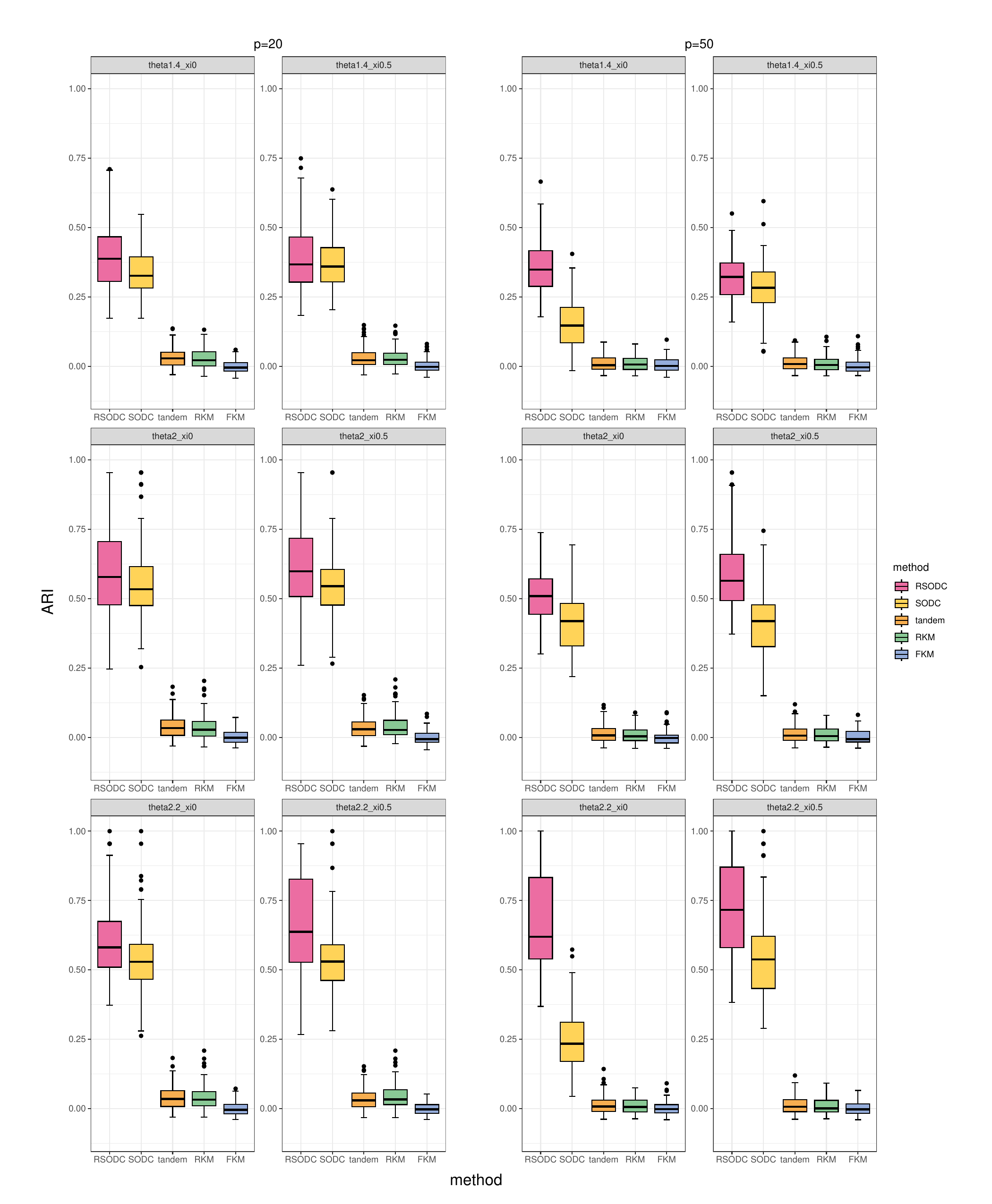}
\caption{Results of ARI in $k=4, n=60, p=20$ and $50$.}
\label{sim_ARIk4n60p2050}
\end{center}
\end{figure}
%
%
\begin{figure}[htbp]
\begin{center}
\includegraphics[scale=0.45]{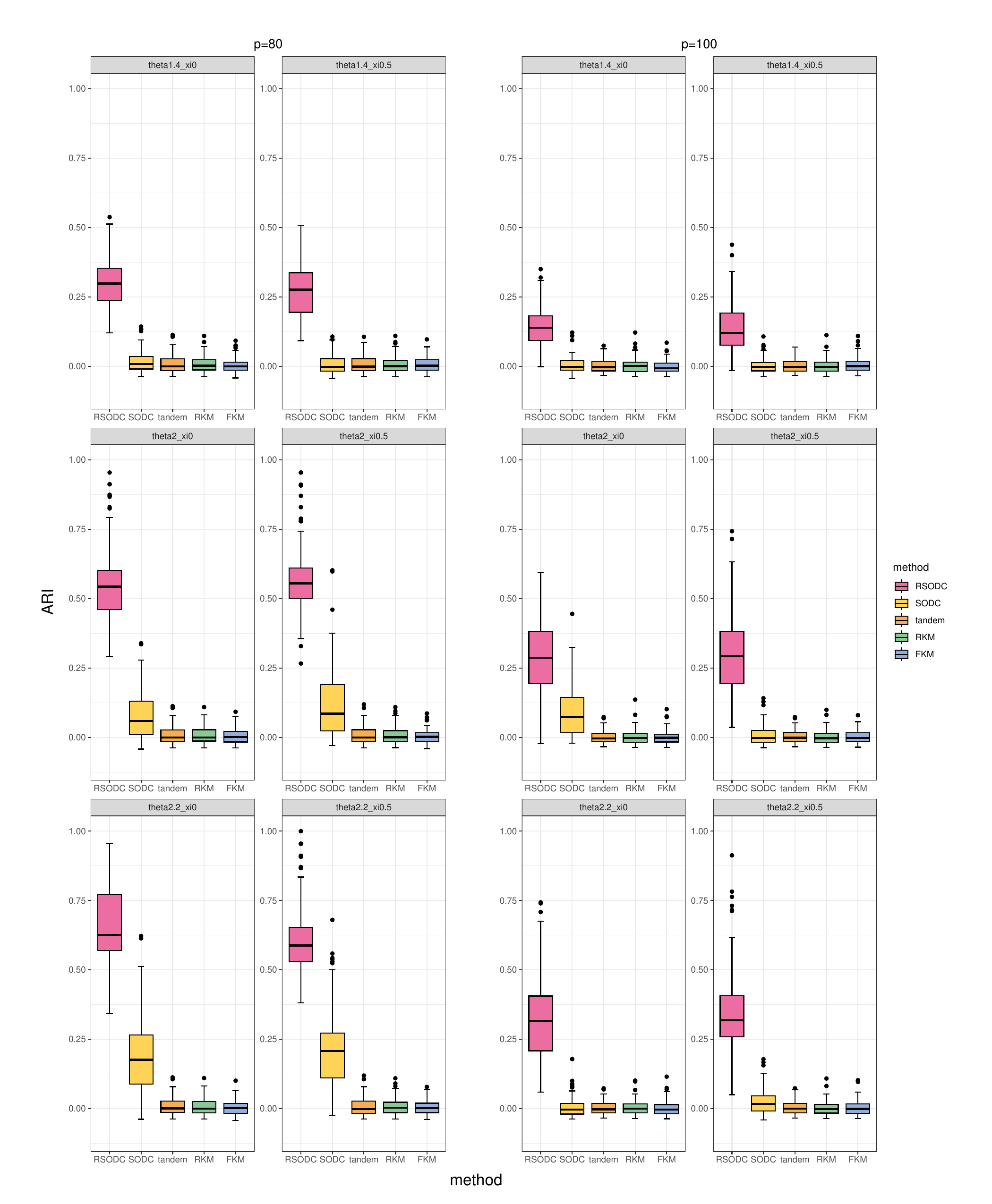}
\caption{Results of ARI in $k=4, n=60, p=80$ and $100$.}
\label{sim_ARIk4n60p80100}
\end{center}
\end{figure}
%

%
\begin{figure}[htbp]
\begin{center}
\includegraphics[scale=0.45]{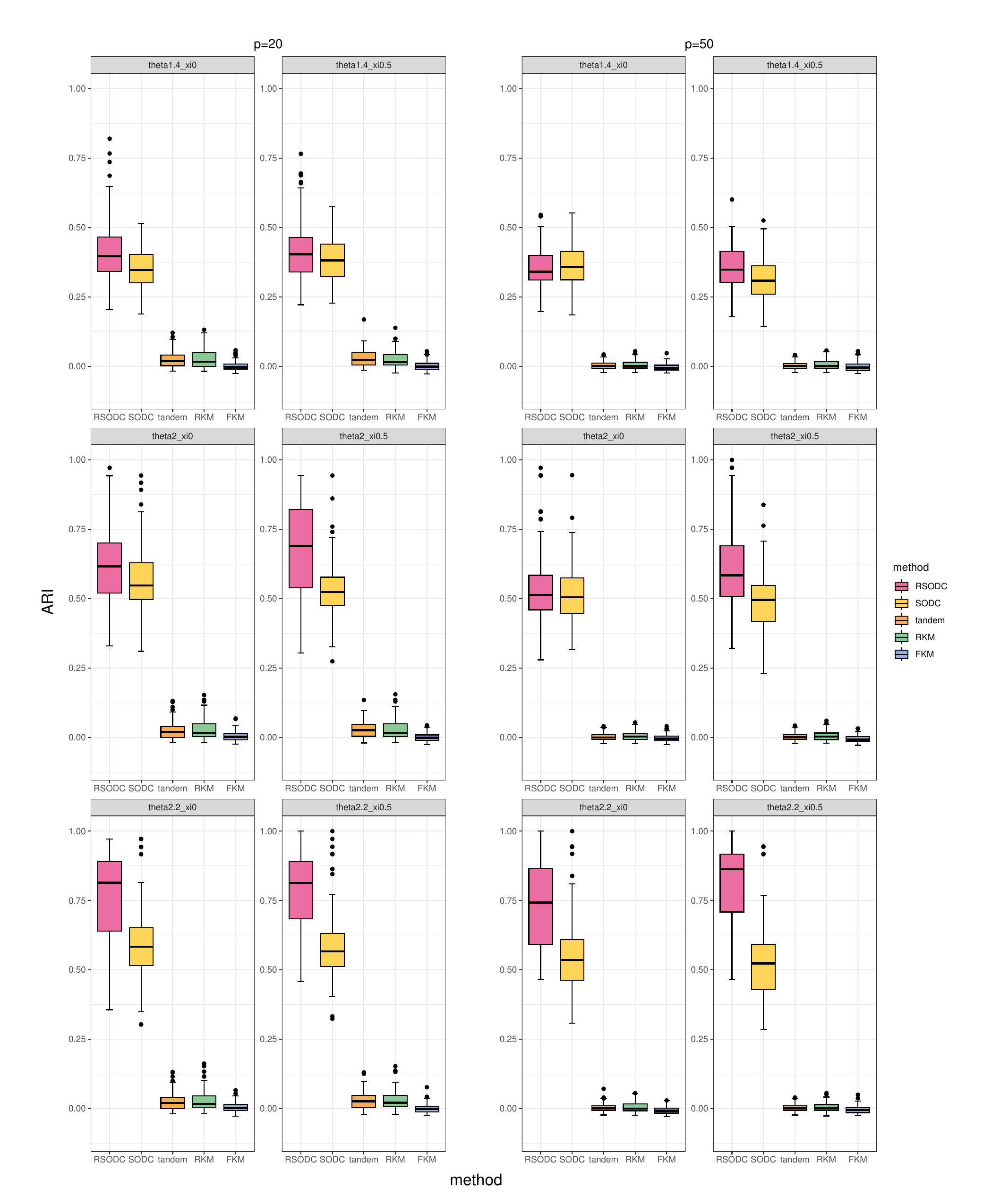}
\caption{Results of ARI in $k=4, n=96, p=20$ and $50$.}
\label{sim_ARIk4n96p2050}
\end{center}
\end{figure}
%
%
\begin{figure}[htbp]
\begin{center}
\includegraphics[scale=0.45]{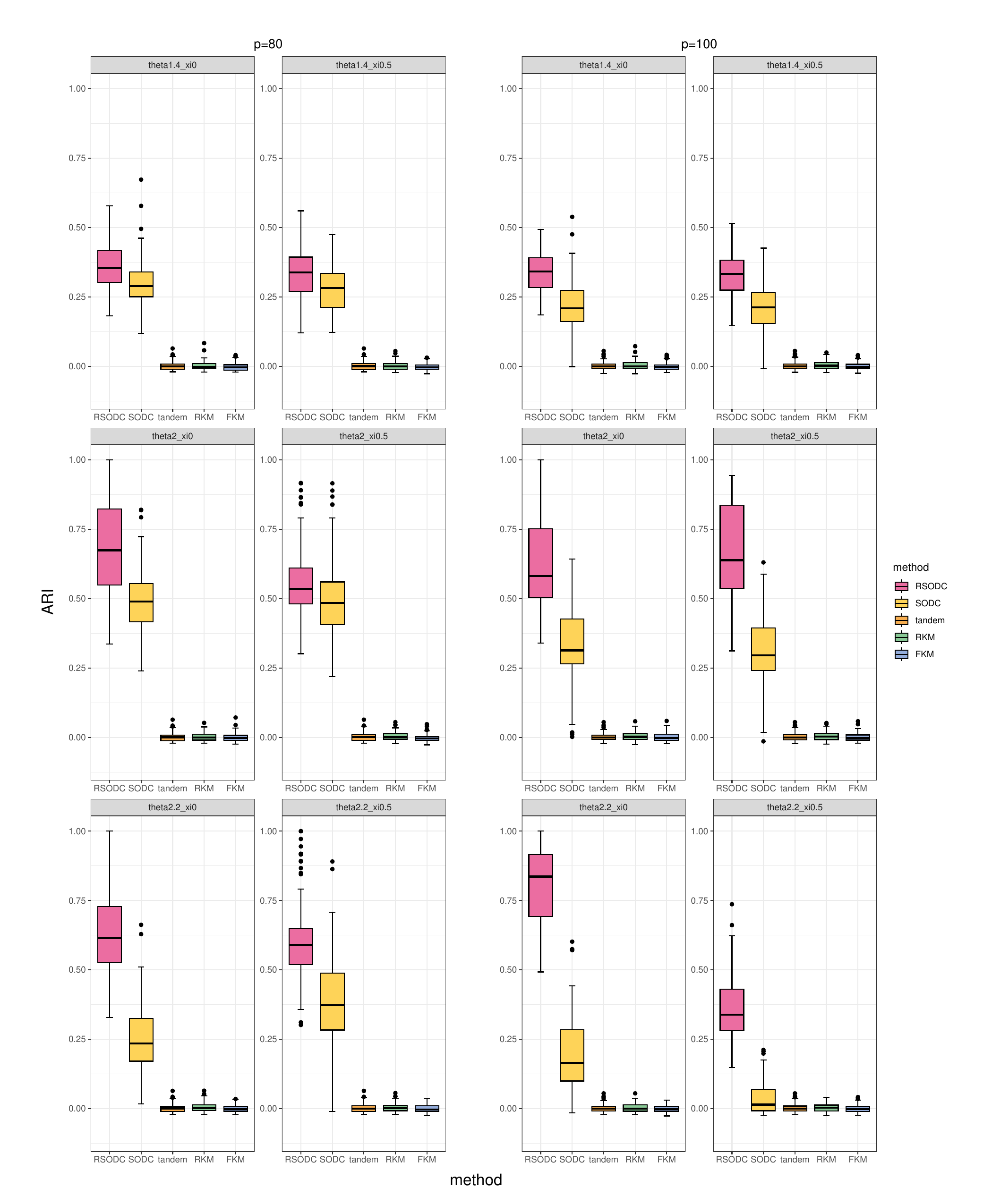}
\caption{Results of ARI in $k=4, n=96, p=80$ and $100$.}
\label{sim_ARIk4n96p80100}
\end{center}
\end{figure}
%

%
\begin{figure}[htbp]
\begin{center}
\includegraphics[scale=0.45]{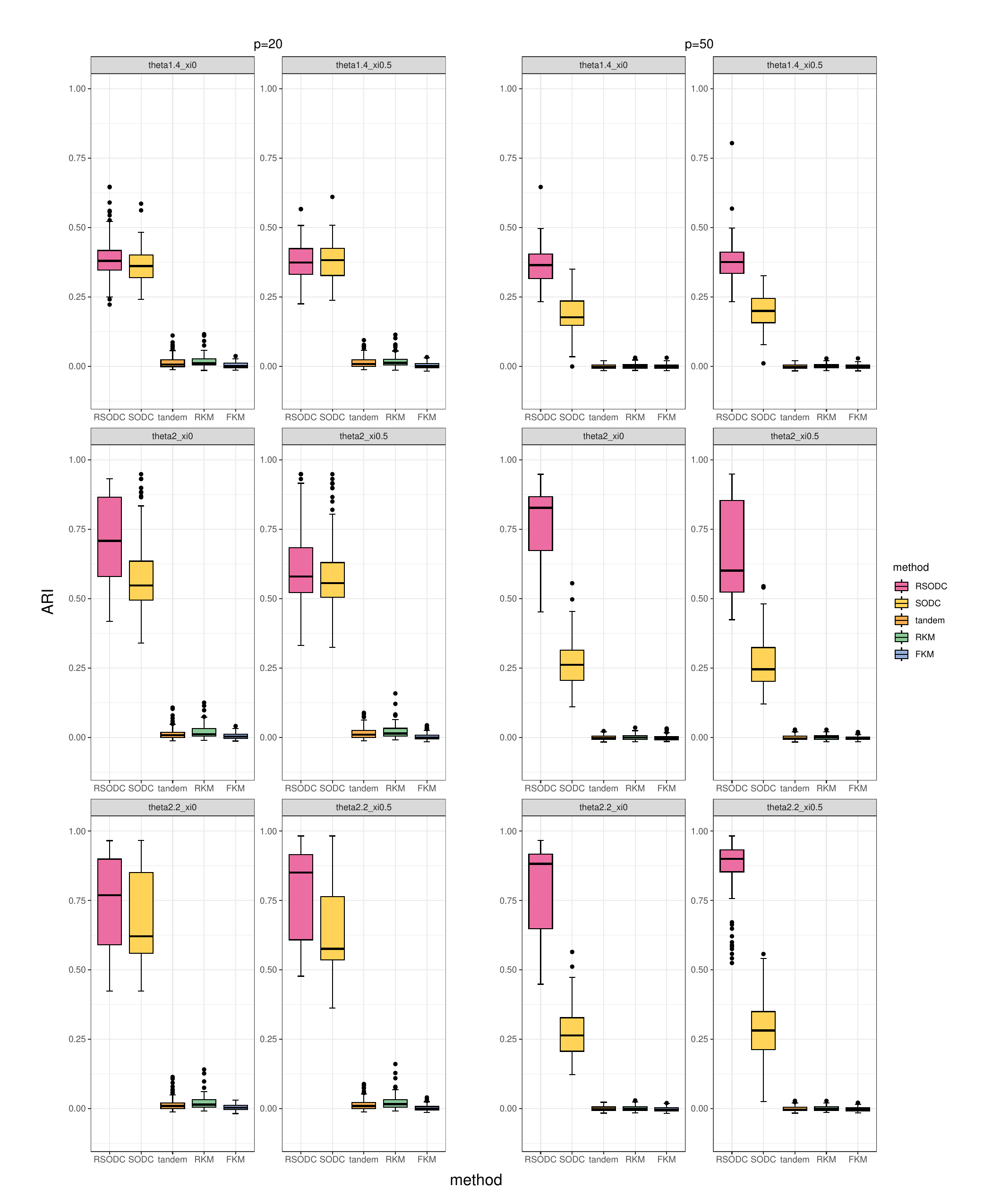}
\caption{Results of ARI in $k=4, n=156, p=20$ and $50$.}
\label{sim_ARIk4n156p2050}
\end{center}
\end{figure}
%
%
\begin{figure}[htbp]
\begin{center}
\includegraphics[scale=0.45]{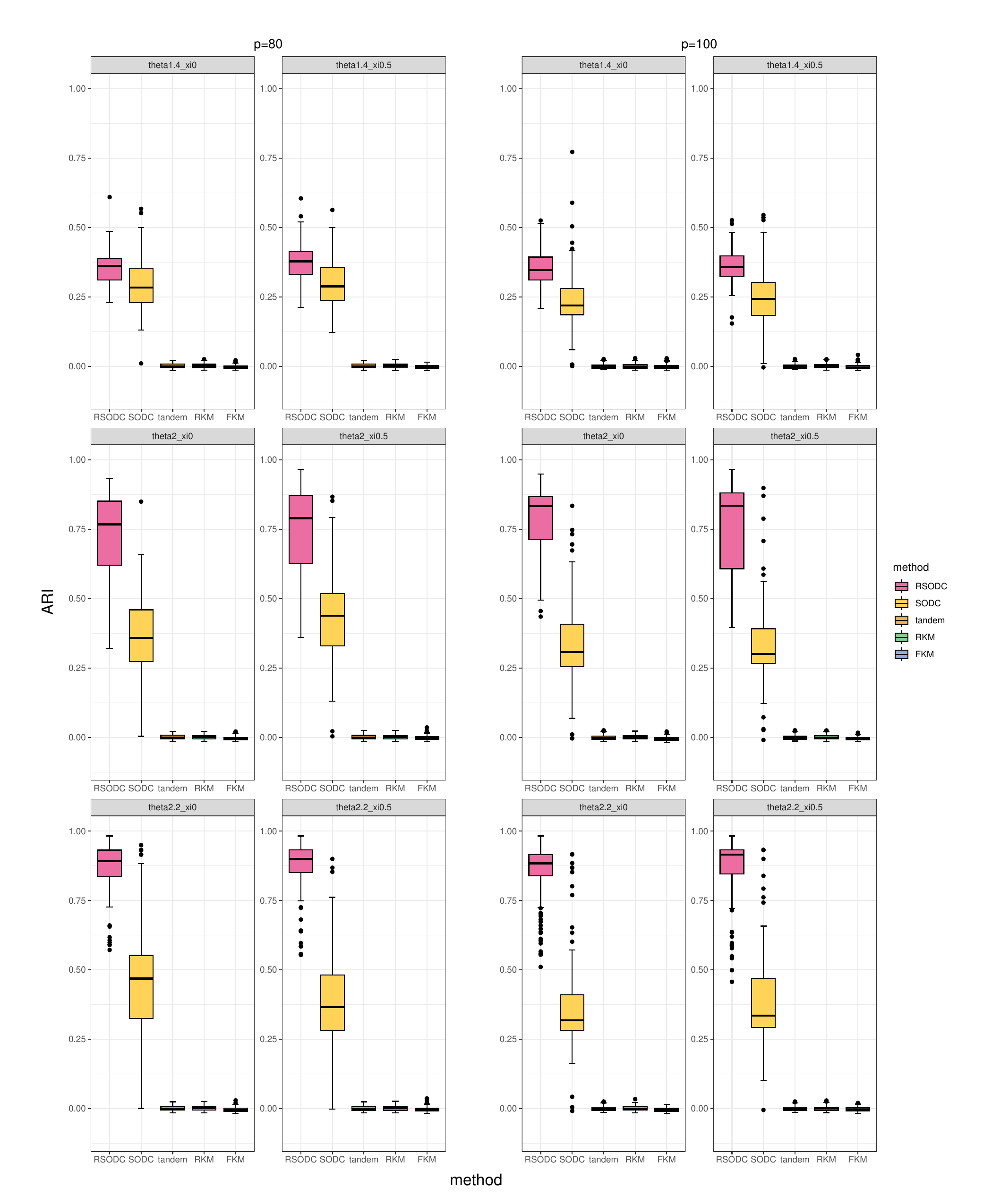}
\caption{Results of ARI in $k=4, n=156, p=80$ and $100$.}
\label{sim_ARIk4n156p80100}
\end{center}
\end{figure}

First of all, the results of Simulation 1 are shown in Figure \ref{sim_ARIk3n60p2050} to Figure \ref{sim_ARIk4n156p80100}. 
The results for $k=3$ are shown in Figure \ref{sim_ARIk3n60p2050} and Figure \ref{sim_ARIk3n156p80100}: Figure \ref{sim_ARIk3n60p2050} and Figure \ref{sim_ARIk3n60p80100} for $n=60$, Figure \ref{sim_ARIk3n96p2050} and Figure \ref{sim_ARIk3n96p80100} for $n=96$, and Figure \ref{sim_ARIk3n156p2050} and Figure \ref{sim_ARIk3n156p80100} for $n=156$. 
In $k=3$, RSODC demonstrated better ARI results for almost all patterns than all compared methods. 
We consider the results by each factor. For $n$, no significant trend appeared in all methods. 
For $p$, the proposed method remained stable while increasing $p$. However, the ARI of SODC became unstable as $p$ increased, especially in $(n,p) = (60, 80), (60,100), (156,100)$. In these situations, the range of ARI was wide.  
For $\vartheta$, the distance between cluster centroids, the ARI values of RSODC and SODC increased as $\vartheta$ became larger. This trend was more noticeable in the proposed method. 
As the value of $\xi$ increased, the ARI value increased in both RSDOC and SODC.

Next, the results for $k=4$ are presented. The results for $k=4$ and $n=60$ are shown in Figure \ref{sim_ARIk4n60p2050} and Figure \ref{sim_ARIk4n60p80100}, those for $n=96$ are in Figure \ref{sim_ARIk4n96p2050} and Figure \ref{sim_ARIk4n96p80100}, and those for $n=156$ are in Figure \ref{sim_ARIk4n156p2050} and Figure \ref{sim_ARIk4n156p80100}. 
The proposed method performed better than the other compared methods for all patterns; SODC performed better, followed by the proposed method. 
First, as for the results for $n$, it did not have a particular influence on the results of all methods. 
However, the number of covariates $p$ also did not affect the results in RSODC. On the other hand, SODC was more unstable when increasing $p$, which is similar to the results in $k=3$. Observing the results for each $\vartheta$, the value of ARI has increased as the value of $\vartheta$ increased for RSODC. While this is the same trend as for $k=3$, the degree of increase for the proposed method was greater for $k=4$. In SODC, this trend was generally the same as for RSODC, however, as $\vartheta$ increased, the ARI values occasionally deteriorated, especially when $p$ was large.  
The change due to $\xi$ was less than for the case of $k=3$ except $n=96$ and $p=100$. 
In addition to ARI, we show the results of the computation time between RSODC and SODC. 
The calculation time for both RSODC and SODC generally increased depending on the number of $n$ and $p$. The results are in Table \ref{sim1_calc} in Appendix \ref{secA1}.

Then, we explain the results of Simulation 2. The median and the mean of ARI are shown in Table \ref{sim2_median} and Table \ref{sim2_mean} in Appendix \ref{secA1}, respectively. Both results show that ARI did not differ significantly among the patterns of the tuning parameters in this simulation setting. 

In Simulation 3, Table \ref{sim3_table} in Appendix \ref{secA1} describes the results. In both cluster settings, the true number of the clusters was selected the most in this simulation. 
In this setting, gap statistics are considered effective for determining the number of clusters.

After that, we observe the results of Simulation 4. Figure \ref{sim4_ari}, Table \ref{sim4_calc}, and Table \ref{sim4_sens} in Appendix \ref{secA1} are the results of Simulation 4. 
Figure \ref{sim4_ari} describes the ARI plotted by $\delta$. 
In $\delta=0.01$ and $\tau=0$, the median of ARI was highest among all patterns. 
This suggests that the ARI results were favorable when the same weight was applied to $\bm{v}_l$. 
However, the range of values for $\tau=0$ was wider when $\delta$ was larger. 
However, when $\tau=55$, the situation where the nearest neighbors were large enough compared to $n=60$, the median ARI was worse for $\delta = 0.001, 0.005$, and $0.01$. The range of ARI was also wide when $\tau=40, 50$, and $55$ in $\delta=0.05$. 
Next, Table \ref{sim4_calc} shows the median of calculation time and that of the number of convergences. When $\tau=0$ and $\delta = 0.001, 0.005$, and $0.01$, the calculation time was longer by three times longer than for the other patterns. This is due to attributing equal weight to every row vector of $\bm{V}$. 
Thirdly, the differences in $\alpha_l$ did not affect the number of convergences in this setting. 
The results for sensitivity and specificity are shown in Table \ref{sim4_sens}. As for the sensitivity, RSODC estimated the two informative variables as non-zero in almost all patterns. For specificity, the pattern $\delta=0.01$ and $\tau=0$ estimated all non-informative variables as zero. The patterns $\delta=0.001$ and $0.005$ also had a high rate for $\tau=0$.

Finally, the results of Simulation 5 are summarized in Table \ref{sim5_summary} in Appendix \ref{secA1}. 
The ARI results tend to follow a similar pattern to those in Simulation 1. 
There was no significant difference in the calculation time and the number of convergences depending on the initial value of $\bm{B}$ in this simulation setting.

\section{Real data application}
\label{sec:realdata}

\subsection{Data description \label{subsec:data_socdc}}

For the application to real data, we employ the proteomics data on breast cancer named "breast TCGA" \citep{real_protein} from R package {\choosefont{pcr}mixOmics}  \citep{mixOmics}. This data contains $142$ different proteins from $150$ subjects, with the cancer subtype being $3$; Basal, Her2, and LumA. 
The evaluation in numerical simulations was conducted on data with uniform cluster sizes, whereas in real data, the performance evaluation is performed on data with non-even cluster sizes. 
This case also aims to assess the performance in the scenario where the data contains a smaller number of more informative variables, as well as to evaluate the sensitivity and specificity of the estimated $\bm{B}$. 
Therefore, we select $10$ informative variables and $70$ less informative variables based on F-values \citep{F-value}. 
The F-value is calculated from the variance ratio in the analysis of variance (ANOVA); a higher F-value indicates a greater variance between each class, suggesting a stronger ability to identify class structure. Conversely, a lower F-value indicates that the variable does not contribute significantly to class structure, and thus is treated as less informative variables.

The evaluation indices includes Adjusted Rand Index (ARI) between the estimated clustering structure and the true clustering structure, variance ratio in estimated $\bm{B}$ and $\bm{Y}$, and the sensitivity and specificity of $\bm{B}$. 
The variance ratio is calculated by the ratio between the variance within clusters and that between clusters. A higher variance ratio signifies that the clusters are more identified. 

The sensitivity and specificity are calculated as follows:\\ 

\begin{align*}
Sensitivity  = \frac{{\rm the \ number \ of \ nonzero \ elements \ corresponding \ to \ informative \ variables }}{10 \times (k-1)},
\end{align*}

\begin{align*}
Specificity  = \frac{{\rm the \ number \ of \ zero \ elements \ corresponding \ to \ less \ informative \ variables }}{70 \times (k-1)}.
\end{align*}

The parameters of $\eta_1$, $\gamma$, and $\rho$ are determined through cross-validation based on kappa statistics, as in the numerical simulations, and $\eta_2$ in RSODC and SODC is set to $0$. 
For the compared methods, five methods are applied for ARI: SODC, tandem clustering in $(k-1)$ dimension, reduced $k$-means in $(k-1)$ dimension, factorial $k$-means in $(k-1)$ dimension, and t-SNE \citep{tsne}. 
For variance ratio,  we compare the proposed method with SODC, tandem-clustering, and t-NSE, while the sensitivity and specificity are compared between RSODC and SODC.

\subsection{Results of data application \label{subsec:relres_socdc}}

\begin{figure}[htbp]
\begin{center}
\includegraphics[scale=0.55]{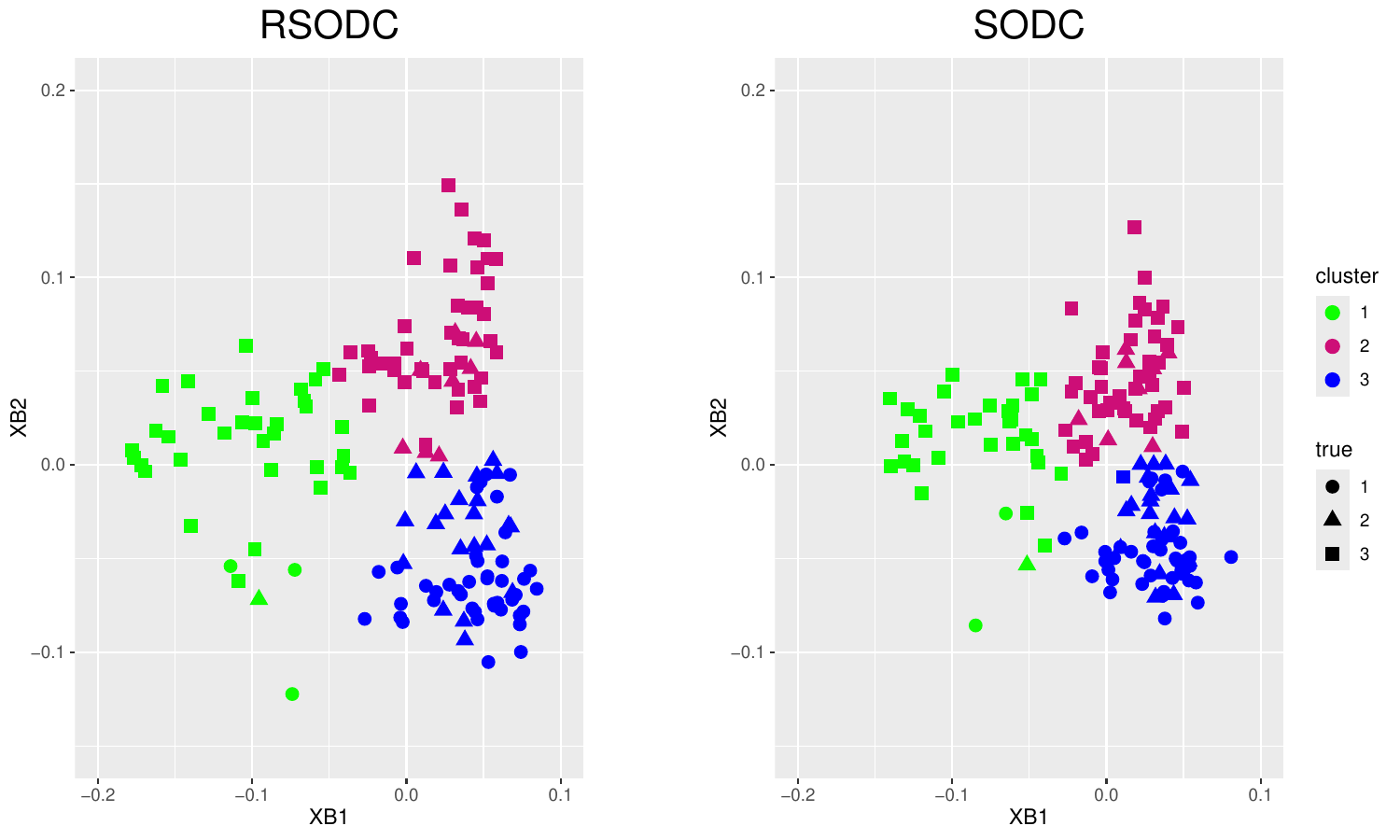}
\caption{Result of $\bm{X}\hat{\bm{B}}$ in RSODC and SODC. The horizontal axis represents the first column of $\bm{X} \hat{\bm{B}}$ and the vertical axis represents the second column. The color of the points indicates the clusters estimated by the methods, and the shape of the points indicates the true clusters.}
\label{plot_xw_pro}
\end{center}
\end{figure}
 
\begin{figure}[htbp]
\begin{center}
\includegraphics[scale=0.55]{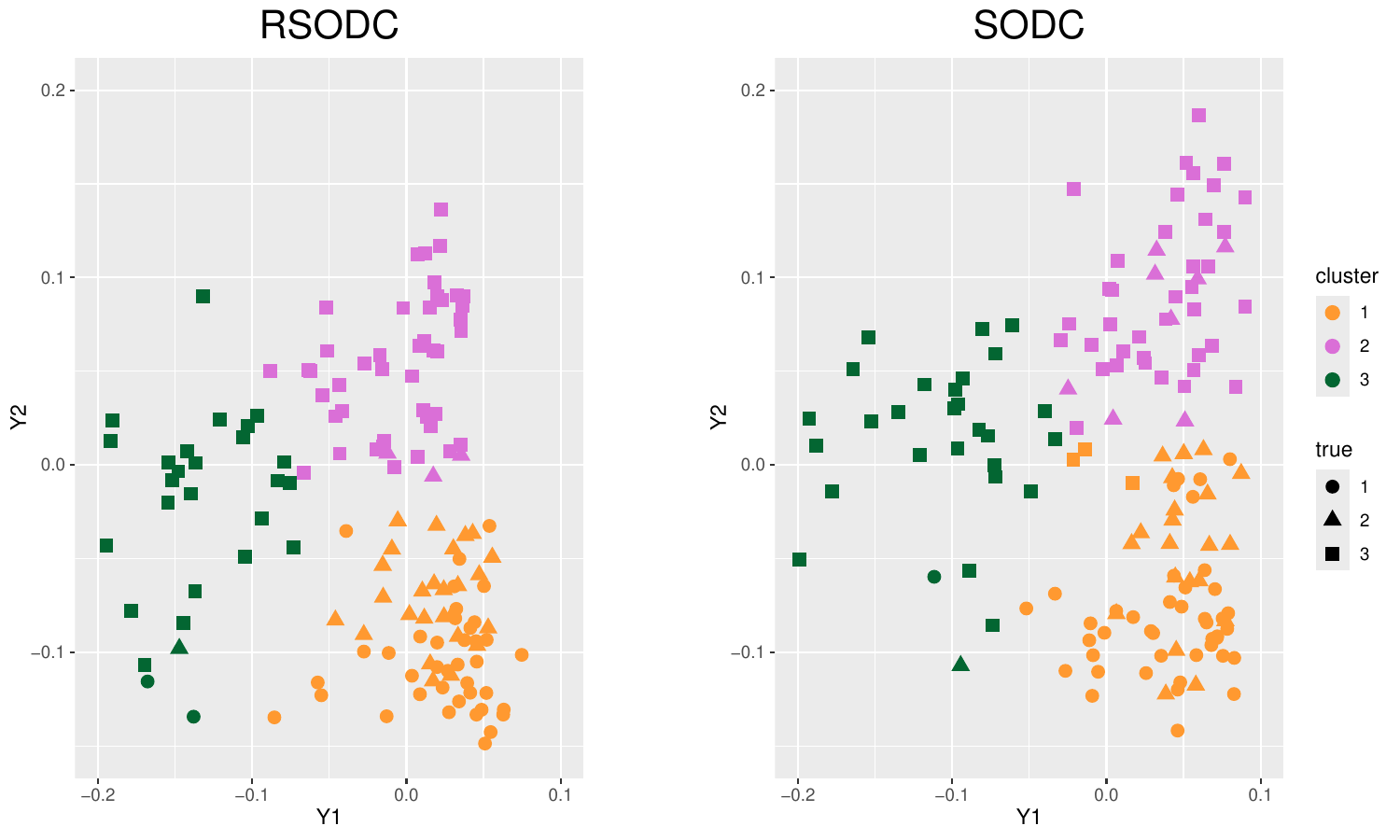}
\caption{Result of $\bm{\hat{Y}}$ of the proposed method and SODC. The horizontal axis represents the first column of $\bm{\hat{Y}}$ and the vertical axis represents the second column. The color of the points indicates the clusters estimated by the methods, and the shape of the points indicates the true clusters.}
\label{plot_est_y_pro}
\end{center}
\end{figure}

\begin{table}[h]
\centering
\caption{Results of ARI. "tandem" indicates tandem clustering, "RKM" indicates reduced $k$-means, and "FKM" denotes factorial $k$-means. In RSODC and SODC, ARI is computed using the clustering results obtained by applying $k$-means to the estimated $\bm{B}$ and $\bm{Y}$ respectively.}
\label{real_ari_pro}
\begin{tabular}{crrrrrrrr} 
\toprule
 & \multicolumn{2}{c}{$\bm{X\hat{B}}$} & \multicolumn{2}{c}{$\bm{\hat{Y}}$} &&&\\
& RSODC & SODC & RSODC & SODC & tandem & t-SNE & RKM & FKM \\ \hline
ARI & $\bm{0.406}$ & $0.401$  & $\bm{0.441}$ & $0.363$ & $0.397$ & $0.285$ & $-0.018$ & $-0.013$ \\\hline
\end{tabular}
\end{table}

\begin{table}[h]
\centering
\caption{Results of variance ratio in $\bm{X\hat{B}}$ and $\bm{\hat{Y}}$ of RSODC and SODC, tandem clustering, and that of t-SNE.}\label{real_var_pro}
\begin{tabular}{crrrrrr} 
\toprule
 & \multicolumn{2}{c}{$\bm{X\hat{B}}$} & \multicolumn{2}{c}{$\bm{\hat{Y}}$} &\\
 & RSODC & SODC  & RSODC & SODC & tandem & t-SNE\\ \hline
variance ratio & $\bm{3.038}$ & $2.909$ & $\bm{3.056}$ & $2.660$ & $2.739$ & $3.033$\\\hline
\end{tabular}
\end{table}

\begin{table}
\centering
\caption{Results of sensitivity and specificity of the proposed method and SODC.}\label{real_sens_spec_pro}
\begin{tabular}{crrrr} 
\toprule
 & \multicolumn{2}{c}{sensitivity} & \multicolumn{2}{c}{specificity}\\
 & RSODC & SODC  & RSODC & SODC \\ \hline
 & \bm{$0.100$} & \bm{$0.100$} & $0.857$ & $\bm{0.971}$ \\\hline
\end{tabular}
\end{table}

The results of ARI are described in Table \ref{real_ari_pro}. The values of the proposed method were superior to those of all compared methods. In terms of $\bm{\hat{Y}}$, the ARI value of RSODC was better than that for $\bm{X\hat{B}}$. The ARI value of SODC in $\bm{\hat{Y}}$ was less than that in $\bm{X\hat{B}}$. 
This is thought to reflect that the $\bm{\hat{Y}}$ obtained from the RSODC, having been given a clustering structure by $\bm{V}$, enabled this clustering structure to be captured more clearly. 

Next, the results of variance ratio are shown in Table \ref{real_var_pro}. 
The values of RSODC in both $\bm{X\hat{B}}$ and $\bm{\hat{Y}}$ were larger than those of the compared methods. 
This indicates that RSODC could be estimated such that points within the same cluster are closer together, while points between clusters are more separated. 
The plots of $\bm{X\hat{B}}$ and $\bm{\hat{Y}}$ in RSODC and SODC are shown in Fig. \ref{plot_xw_pro} and Fig. \ref{plot_est_y_pro}, respectively. 
In RSODC, data points tended to be plotted further apart between different clusters. 
Particularly, in Fig. \ref{plot_est_y_pro}, compared to SODC, data points in RSODC are more clearly separated between each cluster, and those within the same cluster are plotted more densely compared to SODC. 
These trends align with the results of the variance ratio. 
In addition, compared to the plots of tandem clustering and t-SNE shown in Fig. \ref{plot_rel_tand_tsne} in Appendix \ref{secA2}, which depict the results of applying the principal components for tandem clustering and the coordinates obtained by t-SNE to $k$-means, this trend is more apparent. 

Finally, the results show the sensitivity and specificity of $\bm{B}$ in RSODC and SODC in Table \ref{real_sens_spec_pro}. In this application, no difference in sensitivity was observed between these two methods. However, SODC exhibited higher specificity.
That is, SODC tended to estimate more unnecessary variables as zero, resulting in more sparse estimated results. Conversely, RSODC tended to leave some unnecessary variables non-zero.

\section{Discussion and Conclusion}
\label{sec:concl}

In this study, we proposed a novel SODC method with the penalty term derived from convex clustering applied to clustering the scoring matrix $\bm{Y}^\dagger$. 
By adding this penalty term, RSODC was able to capture the clustering structure more clearly compared to SODC.  
The results demonstrated that RSODC brought points from the same cluster closer together and separated the different points more effectively than SODC, as shown by the variance ratio in real data application. 
Throughout the numerical simulations, it was found that the performance of clustering identification was also compared favorably with other dimension reduction clustering methods in this setting. 
We also developed an algorithm by using the majorizing function to derive the updated formula of $\bm{Y}$. This enabled satisfying the orthogonal constraint on $\bm{Y}$ and containing the clustering structure simultaneously. 
The proposed method uses orthogonal Procrustes analysis to update $\bm{Y}$. However, the terms related to $\bm{Y}$ consisted of both quadratic and linear forms, whereas it should only be expressed in linear form. 
Consequently, we derived a majorizing function to ensure that it could be represented exclusively in linear form.

The results of the numerical simulations are discussed in detail. RSODC performed better overall in the setting of this study than the compared methods in Simulation 1. This shows that RSODC was effective in capturing the clustering structures and maintaining estimation accuracy when the number of true clusters was small. 
However, the calculation time of SODC was shorter than that of RSODC, especially when $n$ was larger. 
RSODC incurs higher computation costs due to differences between each row of $n$, and more parameters need to be estimated using ADMM compared to SODC. 
However, the estimation of SODC became unstable as $p$ increased. From these results, RSODC is considered to have received more stability with fewer clustering information compared to SODC by assuming that a clustering structure exists and adds a regularization term to $\bm{Y}^\dagger$. In this study, since the Ridge term was calculated as $0$, it is considered that the addition of a Ridge penalty term \citep{hoerl_ridge1970} could enhance the stability of the estimation in both RSODC and in SODC. 

Next, regarding the results of selecting the number of clusters, in Simulation 3, the true number of clusters was selected most frequently when determining the number of clusters using gap statistics in this simulation setting. 
Determining the number of clusters using gap statistics is considered to be reasonably effective in this proposed method. 
Meanwhile, $n$ was small in this setting, so there is potential for improvement when $n$ is larger.

The selection of the parameters of the weight $\alpha_l$ for the regularization $\bm{V}$ was found to influence the estimation in RSODC through Simulation 4. 
The median ARI value was highest when equal weights were applied to the differences in $\bm{Y}$. 
The following factors may have contributed to why the nearest neighbor setting for $\alpha_l$ did not work effectively for estimation. 
In the original convex clustering, the original space possesses a cluster structure; therefore, calculating neighbors is meaningful. 
In contrast, with RSODC, clustering is performed in a low-dimensional space, yet weights are calculated in the original space; consequently, the weights may not have functioned effectively.
Calculating weights in low-dimensional spaces is desirable; however, this is computationally challenging and remains future work.

In real data application, the ARI value demonstrated that the proposed method captured the clustering structure more accurately than the compared methods. 
However, RSODC does not impose assumptions regarding the distribution of specific data, although it implicitly assumes that the error follows a normal distribution due to the use of the least squares criterion. 
Genetic data often do not follow a normal distribution, and therefore, the data distribution is not optimal for the proposed method. This is considered to have influenced on the ARI evaluation.
From the variance ratio, the proposed method was found to improve the estimation by placing different clusters further apart and the same clusters closer together compared to SODC. 
Regarding the estimation of $\bm{B}$, the proposed method estimated the informative variables as non-zero, similarly to SODC. However, SODC estimated non-informative variables to be more sparse than the proposed method. 
In this study, RSODC was applied to genetic data; however, as it is also applicable to social science and survey data, further considerations are required.

There are several aspects that should be considered to enhance RSODC.  
First, cross-validation for parameter $\eta_1$, $\gamma$, and $\rho$ was performed based on the idea of clustering stability, as in the selection method of conventional SODC; however, this method is computationally expensive. 
For $n=60, p=20, \vartheta=1.4, \xi=0$, and $10$ times data splitting in one pattern of the tuning parameters, the calculation cost summed to $11$ minutes. This will take longer when $n$ and $p$ increase. To reduce computational cost, consideration should be given to accelerating the calculation and deriving an information criterion. 
In addition, in this cross-validation, parameters were explored using kappa coefficient based on the idea of stability of variable selection. However, it is necessary to consider implementing cross-validation based on the stability of clustering results. 
Furthermore, although $\eta_2=0$ was set in this study, it is necessary to consider candidate parameters that include this penalty term.  
This study examined the performance of RSODC with respect to the initial value of $\bm{B}$. However, various approaches are possible for providing initial values, and this requires further investigation. 
Finally, in this study, $k$-means was applied to calculate the clustering results, as $k$-means has been used in SODC among various methods. Comparing the calculation results with other clustering methods should be considered.

\bibliographystyle{unsrtnat}
\bibliography{references}  






\clearpage
\appendix
\appendixpage
\section{Additional Table and Figure in the numerical simulation}\label{secA1}

\begin{table}[htbp]
\centering
\caption{Computation time of RSODC and SODC in Simulation 1. The median of run time is described in second.}\label{sim1_calc}
\footnotesize
\begin{tabular}{ccccrrrrrr} 
\toprule
 & & & & \multicolumn{3}{c}{RSODC} & \multicolumn{3}{c}{SODC} \\
 $k$ & $p$ & $\theta$ & $\xi$ & $n=60$ & $96$  & $156$ & $60$ & $96$  & $156$ \\ \hline
 $3$ & $20$ & $1.4$ & $0$ & $1.081$ &  $9.431$ & $29.651$ &  $0.031$ &  $0.036$ & $0.105$ \\
 & &  & $0.5$ & $1.056$ &  $8.841$ & $29.634$ &  $0.035$ &  $0.041$ & $0.098$ \\
 & & $2.0$ & $0$ & $1.178$ &  $8.077$ & $29.638$ &  $0.019$ &  $0.042$ & $0.092$ \\
 & &  & $0.5$ &  $1.218$ & $8.141$ & $29.555$ &  $0.016$ &  $0.040$ & $0.087$ \\
 & & $2.2$ & $0$ &  $1.118$ &  $8.352$ & $29.210$ &  $0.039$ &  $0.041$ & $0.081$ \\
 &  & & $0.5$ &  $1.243$ &  $7.931$ & $29.837$ &  $0.038$ &  $0.040$ & $0.083$ \\
   &  $50$ & $1.4$ & $0$ & $1.351$ & $11.381$ & $76.128$ & $0.426$ & $0.543$ & $0.855$ \\
  & &  & $0.5$ & $1.391$ &  $9.774$ & $59.767$ &  $0.287$ &  $0.790$ & $0.831$ \\
  & & $2.0$ & $0$ & $1.631$ & $10.572$ & $56.155$ &  $0.617$ & $0.645$ & $0.844$ \\
  & &  & $0.5$ & $1.495$ & $11.669$ & $46.874$ &  $0.315$ &  $0.522$ & $0.766$ \\
  & & $2.2$ & $0$ & $1.341$ & $10.085$ & $46.680$ &  $0.561$ & $0.727$ & $0.728$ \\
  & & & $0.5$ & $1.342$ & $10.021$ & $46.977$ &  $0.389$ & $0.632$ & $0.767$ \\
  & $80$ & $1.4$ & $0$ &  $6.198$ & $15.526$ & $84.488$ & $2.942$ & $1.553$ & $1.120$ \\
  & &  & $0.5$ & $4.238$ & $17.901$ & $86.615$ &  $3.387$ &  $2.691$ & $1.432$ \\
  & & $2.0$ & $0$ & $3.690$ & $16.289$ & $63.931$ &  $0.868$ & $1.688$ & $1.334$ \\
  & &  & $0.5$ & $5.209$ & $15.992$ & $54.557$ &    $1.100$ &  $1.576$ & $1.332$ \\
  & & $2.2$ & $0$ & $5.100$ & $13.001$ & $54.105$ &  $1.977$ &  $1.980$ & $1.381$ \\
  & & & $0.5$ & $5.140$ & $17.604$ & $52.162$ &  $1.622$ & $1.644$ & $1.302$ \\
  & $100$ & $1.4$ & $0$ & $9.368$ & $39.391$ & $111.260$ &  $8.432$ & $9.214$ & $2.735$ \\
  & &  & $0.5$ & $98.397$ & $30.630$ & $95.516$ & $11.990$ & $4.564$ & $2.654$ \\
  & & $2.0$ & $0$ & $75.512$ & $34.023$ & $78.977$ & $10.928$ &  $4.720$ & $2.265$ \\
  & &  & $0.5$ & $20.351$ & $28.073$ & $74.662$ &  $6.912$ & $4.412$ & $2.155$ \\
  & & $2.2$ & $0$ &  $8.642$ & $36.270$ & $47.667$ &  $5.606$ &  $4.140$ & $2.022$ \\
  & & & $0.5$ & $9.746$ & $36.730$ & $42.731$ &  $4.404$ & $14.892$ & $2.102$ \\\hline
  $4$ & $20$ & $1.4$ & $0$ & $1.189$ & $8.960$ &   $49.581$ & $0.037$ & $0.153$ & $0.116$ \\
  & &  & $0.5$ & $1.282$ &  $9.179$ & $55.090$ &  $0.031$ & $0.149$ & $0.112$ \\
  & & $2.0$ & $0$ & $1.233$ &  $9.402$ & $64.723$ &  $0.028$ & $0.147$ & $0.109$ \\
  & &  & $0.5$ & $1.241$ &  $9.358$ & $67.025$ &  $0.031$ & $0.150$ & $0.108$ \\
  & & $2.2$ & $0$ & $1.215$ &  $8.845$ &   $67.524$ &  $0.033$ & $0.140$ & $0.108$ \\
   & & & $0.5$ &  $1.243$ & $8.558$ &  $65.926$ &  $0.034$ & $0.148$ & $0.107$ \\
   & $50$ & $1.4$ & $0$ & $2.835$ & $22.506$ &   $52.852$ & $1.402$ & $1.523$ & $1.232$ \\
  & &  & $0.5$ & $3.402$ & $20.819$ & $52.909$ &   $0.520$ & $1.812$ & $1.151$ \\
  & & $2.0$ & $0$ & $4.292$ & $19.060$ &   $52.857$ &  $0.647$ & $1.693$ & $1.192$ \\
  & &  & $0.5$ & $3.144$ & $16.071$ & $53.337$ &  $0.907$ & $1.726$ & $1.193$ \\
  & & $2.2$ & $0$ & $2.402$ & $15.143$ &   $52.608$ &  $2.553$ & $1.452$ & $1.142$ \\
  & & & $0.5$ & $2.234$ &  $14.234$ &  $49.711$ &  $0.809$ & $1.471$ & $1.115$ \\
   & $80$ & $1.4$ & $0$ & $29.938$ & $28.583$ &  $109.305$ & $31.892$ &  $4.980$ & $5.167$ \\
  & &  & $0.5$ & $49.440$ & $35.699$ & $101.178$ & $51.296$ & $5.026$ & $5.125$ \\
  & & $2.0$ & $0$ & $44.930$ & $37.923$ & $88.482$ & $20.044$ & $3.956$ & $5.043$ \\
  & &  & $0.5$ & $26.976$ & $49.451$ & $84.118$ & $14.536$ & $3.943$ & $5.371$ \\
  & & $2.2$ & $0$ & $38.224$ & $49.612$ &   $79.377$ &  $9.917$ &   $9.860$ & $4.839$ \\
  &  & & $0.5$ & $50.475$ & $56.366$ & $79.094$ &  $9.568$ & $8.419$ &  $5.132$ \\
  & $100$ & $1.4$ & $0$ & $291.015$ & $100.126$ &  $216.433$ & $33.639$ & $12.475$ & $10.062$ \\
 & &  & $0.5$ & $328.926$ &  $80.000$ & $145.116$ & $77.718$ & $9.913$ & $9.652$ \\
  & & $2.0$ & $0$ & $320.594$ &  $72.797$ &  $143.663$ & $12.024$ & $9.690$ & $9.797$ \\
  & &  & $0.5$ & $361.026$ & $61.242$ & $128.415$ & $36.015$ & $9.242$ & $9.482$ \\
 &  & $2.2$ & $0$ & $354.030$ &  $57.058$ &  $124.991$ & $36.405$ & $21.333$ & $10.888$ \\
 &  & & $0.5$ & $347.534$ & $440.968$ & $126.436$ & $26.148$ & $46.885$ & $10.907$ \\\hline
\end{tabular}%
\end{table}

\newpage

\begin{table}
\centering
\caption{Median of ARI in Simulation 2 by $\eta_1$ in Simulation 2.}\label{sim2_median}
\begin{tabular}{ll|cccccccc}
\hline
& $\eta_1$ & 0.1 & 0.5 & 1 & 1.5 & 2 & 2.5 & 3 & 3.5 \\ 
$\gamma$ & $\rho$ & & & & & & &\\ \hline
$0.001$ & $0.01$ & $0.426$ & $0.439$ & $0.439$ & $0.451$ & $0.451$ & $0.479$ & $0.482$ & $0.438$ \\
$0.003$ & $0.01$ & $0.432$ & $0.444$ & $0.426$ & $0.447$ & $0.450$ & $0.453$ & $0.460$ & $0.465$ \\
$0.005$ & $0.01$ &$0.432$ & $0.434$ & $0.442$ & $0.451$ & $0.453$ & $0.458$ & $0.473$ & $0.453$ \\
$0.007$ & $0.01$ &$0.436$ & $0.441$ & $0.453$ & $0.453$ & $0.461$ & $0.454$ & $0.447$ & $0.454$ \\
$0.001$ & $0.03$ & $0.432$ & $0.437$ & $0.441$ & $0.447$ & $0.453$ & $0.461$ & $0.461$ & $0.457$ \\
$0.003$ & $0.03$ & $0.436$ & $0.443$ & $0.445$ & $0.449$ & $0.451$ & $0.451$ & $0.44$ & $0.429$ \\
$0.005$ & $0.03$ &$0.438$ & $0.434$ & $0.439$ & $0.440$ & $0.444$ & $0.454$ & $0.455$ & $0.470$ \\
$0.007$ & $0.03$ & $0.438$ & $0.446$ & $0.457$ & $0.444$ & $0.453$ & $0.460$ & $0.458$ & $0.466$ \\
$0.010$ & $0.03$ &$0.443$ & $0.445$ & $0.454$ & $0.457$ & $0.458$ & $0.462$ & $0.462$ & $0.464$ \\
$0.001$ & $0.05$ & $0.432$ & $0.435$ & $0.437$ & $0.447$ & $0.451$ & $0.449$ & $0.451$ & $0.452$ \\
$0.003$ & $0.05$ &$0.440$ & $0.440$ & $0.442$ & $0.445$ & $0.448$ & $0.446$ & $0.459$ & $0.451$ \\
$0.005$ & $0.05$ &$0.430$ & $0.436$ & $0.441$ & $0.444$ & $0.447$ & $0.448$ & $0.439$ & $0.434$ \\
$0.007$ & $0.05$ &$0.438$ & $0.440$ & $0.439$ & $0.439$ & $0.445$ & $0.459$ & $0.461$ & $0.463$ \\
$0.010$ & $0.05$ &  $0.448$ & $0.444$ & $0.445$ & $0.449$ & $0.447$ & $0.455$ & $0.484$ & $0.485$ \\
$0.001$ & $0.07$ & $0.432$ & $0.436$ & $0.437$ & $0.443$ & $0.445$ & $0.448$ & $0.448$ & $0.452$ \\
$0.003$ & $0.07$ & $0.437$ & $0.439$ & $0.441$ & $0.445$ & $0.449$ & $0.458$ & $0.450$ & $0.455$ \\
$0.005$ & $0.07$ & $0.438$ & $0.447$ & $0.445$ & $0.444$ & $0.447$ & $0.451$ & $0.449$ & $0.441$ \\
$0.007$ & $0.07$ & $0.430$ & $0.434$ & $0.442$ & $0.442$ & $0.445$ & $0.446$ & $0.435$ & $0.430$ \\
$0.010$ & $0.07$ & $0.435$ & $0.439$ & $0.441$ & $0.444$ & $0.452$ & $0.462$ & $0.463$ & $0.47$ \\
$0.001$ & $0.10$ & $0.432$ & $0.435$ & $0.437$ & $0.441$ & $0.443$ & $0.446$ & $0.447$ & $0.444$ \\
$0.003$ & $0.10$ & $0.439$ & $0.439$ & $0.440$ & $0.445$ & $0.449$ & $0.454$ & $0.452$ & $0.456$ \\
$0.005$ & $0.10$ & $0.439$ & $0.441$ & $0.442$ & $0.446$ & $0.454$ & $0.452$ & $0.450$ & $0.452$ \\
$0.007$ & $0.10$ &$0.433$ & $0.443$ & $0.444$ & $0.445$ & $0.447$ & $0.451$ & $0.447$ & $0.440$ \\
$0.010$ &$0.10$ & $0.435$ & $0.43$ & $0.440$ & $0.444$ & $0.445$ & $0.440$ & $0.432$ & $0.434$ \\
\hline
\end{tabular}
\end{table}

\begin{table}
\centering
\caption{Mean of ARI in Simulation 2 by $\eta_1$ in Simulation 2.}\label{sim2_mean}
\begin{tabular}{ll|cccccccc}
\hline
& $\eta_1$ & 0.1 & 0.5 & 1 & 1.5 & 2 & 2.5 & 3 &3.5 \\
$\gamma$ & $\rho$ & & & & & & &\\ \hline
$0.001$ & $0.01$ & $0.444$ & $0.457$ & $0.460$ & $0.479$ & $0.483$ & $0.511$ & $0.508$ & $0.445$ \\
$0.003$ & $0.01$ & $0.457$ & $0.462$ & $0.448$ & $0.471$ & $0.478$ & $0.492$ & $0.495$ & $0.489$ \\
$0.005$ & $0.01$ & $0.440$ & $0.452$ & $0.456$ & $0.468$ & $0.467$ & $0.478$ & $0.476$ & $0.477$ \\
$0.007$ & $0.01$ &  $0.442$ & $0.452$ & $0.459$ & $0.467$ & $0.477$ & $0.490$ & $0.470$ & $0.465$ \\
$0.001$ & $0.03$ &  $0.450$ & $0.456$ & $0.468$ & $0.475$ & $0.482$ & $0.490$ & $0.500$ & $0.501$ \\
$0.003$ & $0.03$ & $0.457$ & $0.462$ & $0.467$ & $0.477$ & $0.466$ & $0.464$ & $0.450$ & $0.412$ \\
$0.005$ & $0.03$ &  $0.450$ & $0.455$ & $0.462$ & $0.459$ & $0.467$ & $0.483$ & $0.484$ & $0.489$ \\
$0.007$ & $0.03$ &  $0.458$ & $0.468$ & $0.470$ & $0.470$ & $0.470$ & $0.489$ & $0.497$ & $0.494$ \\
$0.010$ & $0.03$ & $0.455$ & $0.458$ & $0.468$ & $0.470$ & $0.472$ & $0.482$ & $0.482$ & $0.485$ \\
$0.001$ & $0.05$ & $0.448$ & $0.457$ & $0.462$ & $0.468$ & $0.477$ & $0.484$ & $0.489$ & $0.497$ \\
$0.003$ & $0.05$ & $0.460$ & $0.463$ & $0.467$ & $0.474$ & $0.479$ & $0.483$ & $0.490$ & $0.480$ \\
$0.005$ & $0.05$ &$0.447$ & $0.455$ & $0.461$ & $0.464$ & $0.458$ & $0.461$ & $0.462$ & $0.419$ \\
$0.007$ & $0.05$ & $0.452$ & $0.460$ & $0.462$ & $0.458$ & $0.467$ & $0.481$ & $0.484$ & $0.469$ \\
$0.010$ & $0.05$ & $0.462$ & $0.462$ & $0.469$ & $0.470$ & $0.468$ & $0.489$ & $0.505$ & $0.510$ \\
$0.001$ & $0.07$ & $0.448$ & $0.456$ & $0.460$ & $0.465$ & $0.474$ & $0.483$ & $0.485$ & $0.494$ \\
$0.003$ & $0.07$ & $0.454$ & $0.460$ & $0.467$ & $0.474$ & $0.479$ & $0.488$ & $0.491$ & $0.492$ \\
$0.005$ & $0.07$ & $0.456$ & $0.463$ & $0.468$ & $0.470$ & $0.471$ & $0.472$ & $0.470$ & $0.447$ \\
$0.007$ & $0.07$ & $0.449$ & $0.455$ & $0.459$ & $0.463$ & $0.457$ & $0.463$ & $0.464$ & $0.416$ \\
$0.010$ & $0.07$ & $0.454$ & $0.456$ & $0.465$ & $0.468$ & $0.474$ & $0.495$ & $0.488$ & $0.484$ \\
$0.001$ & $0.10$ & $0.445$ & $0.455$ & $0.459$ & $0.464$ & $0.471$ & $0.480$ & $0.482$ & $0.488$ \\
$0.003$ & $0.10$ & $0.453$ & $0.459$ & $0.466$ & $0.472$ & $0.479$ & $0.488$ & $0.493$ & $0.495$ \\
$0.005$ & $0.10$ & $0.458$ & $0.462$ & $0.466$ & $0.473$ & $0.479$ & $0.485$ & $0.489$ & $0.489$ \\
$0.007$ & $0.10$ & $0.451$ & $0.463$ & $0.463$ & $0.469$ & $0.470$ & $0.474$ & $0.468$ & $0.447$ \\
$0.010$ & $0.10$ & $0.452$ & $0.456$ & $0.457$ & $0.461$ & $0.459$ & $0.464$ & $0.462$ & $0.421$ \\
\hline
\end{tabular}
\end{table}

\begin{table}
\centering
\caption{Total number of the minimum gap statistics for each true cluster in Simulation 3. Each column describes the cluster set for the calculation of RSODC in advance. The row is the true cluster.}\label{sim3_table}
\begin{tabular}{lcccccccc}
\hline
 & $2$ & $3$ & $4$ & $5$ & $6$ & $7$ & $8$ & $9$\\\hline
 $k=2$ & $\bm{28}$ & $15$ & $20$ & $11$ & $5$ & $6$ & $6$ & $1$\\
$k=3$ & $10$ & $\bm{43}$ & $19$ & $15$ & $4$ & $4$ & $2$ & $1$\\
\hline
\end{tabular}
\end{table}

\begin{figure}[htbp]
\begin{center}
\includegraphics[scale=0.53]{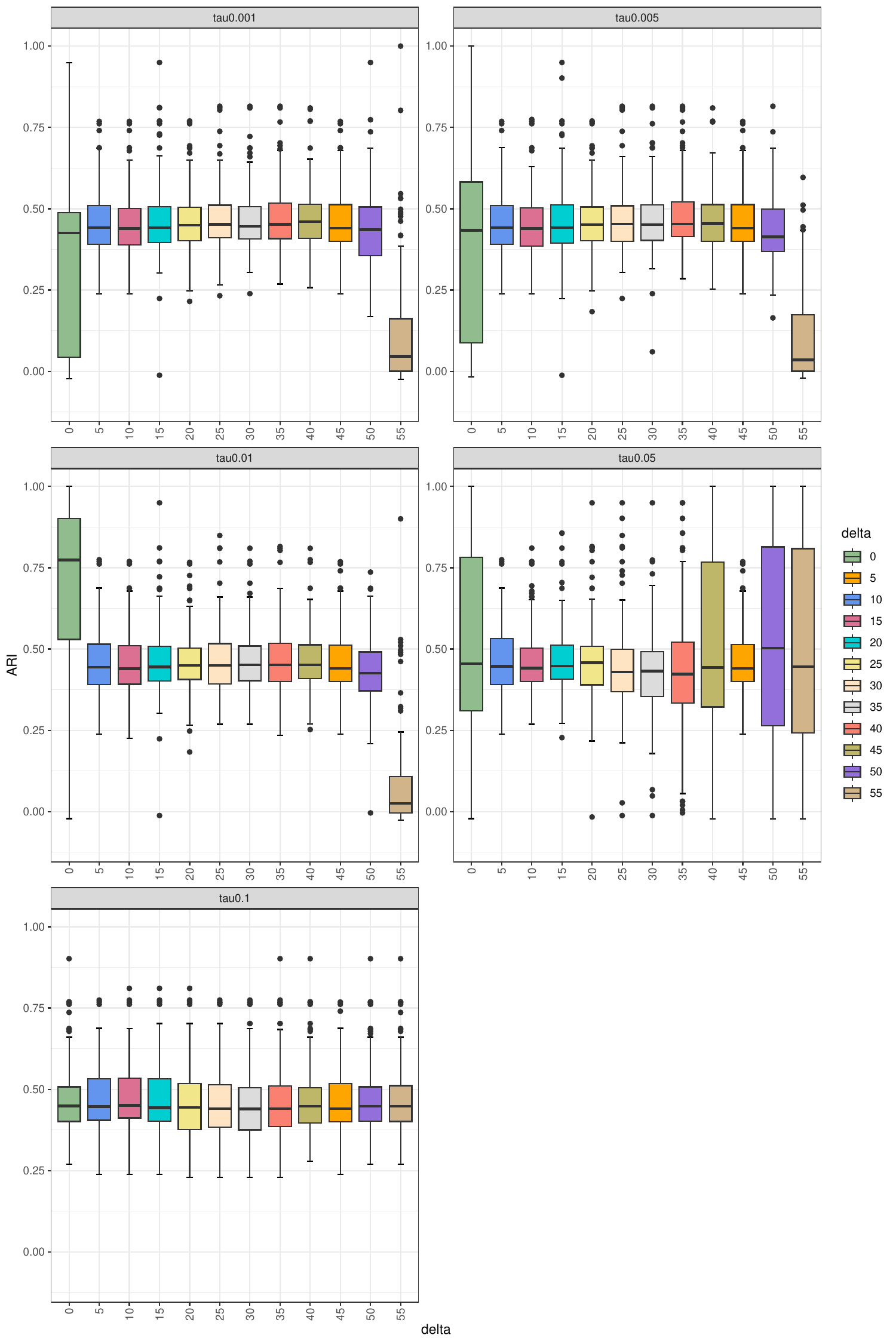}
\caption{Plots of ARI by the weight $\delta$ in Simulation 4. The vertical axis is ARI and the horizontal axis is $\tau$.}
\label{sim4_ari}
\end{center}
\end{figure}
%

\begin{table}[htbp]
\centering
\caption{Median of calculation time and that of the number of convergence of RSODC in Simulation 4. The calculation cost describes by second.}
\label{sim4_calc}
\begin{tabular}{rrrrrrrrrrr}
\toprule
 & \multicolumn{5}{c}{calculation time} & \multicolumn{5}{c}{number of convergence}\\
$\delta$ & $0.001$ & $0.005$ & $0.01$ & $0.05$ & $0.1$ & $0.001$ & $0.005$ & $0.01$ & $0.05$ & $0.1$\\
$\tau$ &  &  &  &  &  &  &  &  &  &\\
\hline
$0$ & $100.493$ & $96.777$ & $96.533$ & $1.059$ & $0.995$ &
$13$ & $13$ & $13$ & $13$ & $13$ \\
$5$ & $0.989$ & $0.989$ & $0.987$ & $0.984$ & $0.992$ & $13$ & $13$ & $13$ & $13$ &$13$ \\
$10$ & $0.989$ & $0.989$ & $0.986$ & $0.981$ & $0.987$ 
& $13$ & $13$ & $13$ &$13$ &$13$ \\
$15$ & $0.988$ & $0.991$ & $0.992$ & $0.991$ & $0.988$ 
&$13$ &$13$ &$13$ &$13$ &$13$ \\
$20$ & $0.995$ & $0.994$ & $0.997$ & $1.067$ & $0.984$ 
&$13$ &$13$ &$13$ &$14$ &$13$ \\
$25$ & $1.000$ & $0.996$ &$0.994$ & $0.999$ & $0.989$ 
&$13$ &$13$ &$13$ & $13$ & $13$\\
$30$ & $1.001$ & $0.998$ & $1.000$ & $1.062$ & $0.992$ & $13$ & $13$ & $13$ & $14$ &$13$ \\
$35$ & $1.059$ & $1.060$ &$1.068$ & $1.166$ & $1.105$ & $13$ & $13$ & $13$ & $13$ &$13$\\
$40$ & $1.148$ & $1.158$ & $1.170$ & $1.103$ & $1.072$ & $13$ & $13$ & $13$ & $13$ & $13$\\
$45$ & $1.544$ & $1.523$ & $1.543$ & $1.059$ & $1.035$ & $13$ & $13$ & $13$ & $13$ &$13$\\
$50$ & $1.617$ & $1.646$ & $1.716$ & $1.083$ & $1.036$ & $13$ & $13$ & $13$ & $13$ & $13$\\
$55$ & $1.696$ & $1.774$ & $1.878$ & $1.184$ & $1.121$ & $13$ & $13$ & $13$ & $13$ & $13$\\
\hline
\end{tabular}
\end{table}
\begin{table}[htbp]
\centering
\caption{Median of sensitivity and specificity of RSODC in Simulation 4.}
\label{sim4_sens}
\begin{tabular}{rcccccccccc}
\hline
 & \multicolumn{5}{c}{sensitivity} & \multicolumn{5}{c}{specificity} \\
$\delta$ & $0.001$ & $0.005$ & $0.01$ & $0.05$ & $0.1$ & $0.001$ & $0.005$ & $0.01$ & $0.05$ & $0.1$ \\
$\tau$ & & & & & & & & & & \\
\hline
$0$ & $0.50$ & $1.00$ & $1.00$ & $1.00$ & $1.00$ & $0.94$ & $0.83$ & $1.00$ & $0.61$ & $0.44$ \\
$5$ & $1.00$ & $1.00$ & $1.00$ & $1.00$ &$1.00$ & $0.44$ & $0.44$ & $0.44$ & $0.44$ & $0.44$ \\
$10$ & $1.00$ & $1.00$ & $1.00$ & $1.00$ & $1.00$ & $0.44$ & $0.44$ & $0.44$ & $0.44$ & $0.44$ \\
$15$ & $1.00$ & $1.00$ & $1.00$ & $1.00$ & $1.00$ & $0.44$ & $0.44$ & $0.44$ & $0.44$ & $0.44$ \\
$20$ & $1.00$ & $1.00$ & $1.00$ & $1.00$ & $1.00$ & $0.44$ &$0.44$ & $0.44$ & $0.44$ &$0.44$ \\
$25$ & $1.00$ & $1.00$ & $1.00$ & $1.00$ & $1.00$ & $0.44$ &$0.44$ & $0.44$ & $0.44$ & $0.44$ \\
$30$ & $1.00$ & $1.00$ & $1.00$ & $1.00$ & $1.00$ & $0.39$ & $0.44$ & $0.39$ & $0.44$ & $0.44$ \\
$35$ & $1.00$ & $1.00$ & $1.00$ & $1.00$ & $1.00$ & $0.39$ & $0.42$ & $0.44$ & $0.44$ & $0.44$ \\
$40$ & $1.00$ & $1.00$ & $1.00$ & $1.00$ & $1.00$ & $0.44$ & $0.44$ &$0.44$ & $0.44$ & $0.44$ \\
$45$ & $1.00$ & $1.00$ & $1.00$ & $1.00$ & $1.00$ & $0.44$ & $0.44$ & $0.44$ & $0.50$ & $0.44$ \\
$50$ & $1.00$ & $1.00$ & $1.00$ & $1.00$ & $1.00$ & $0.50$ & $0.50$ & $0.50$ & $0.56$ & $0.44$ \\
$55$ & $1.00$ & $1.00$ & $0.50$ & $1.00$ & $1.00$ & $0.61$ & $0.61$ &$0.64$ & $0.56$ & $0.44$ \\
\hline
\end{tabular}
\end{table}
%

\begin{table}[htbp]
\centering
\caption{Summary of Simulation 5. "calc\_time" defers to the calculation time of RSODC, and "num\_convergence" is the number of convergence of RSODC.}
\label{sim5_summary}
\begin{tabular}{lccc}
\hline
Summary & ARI & calc\_time & num\_convergence \\
\hline
Min. & $0.226$ &$1.115$ & $13.0$ \\
1st\ Qu. & $0.403$ & $1.121$ & $13.0$ \\
Median & $0.451$ & $1.124$ & $13.0$ \\
Mean & $0.466$ & $1.144$ & $13.2$ \\
3rd\ Qu. & $0.518$ & $1.133$ & $13.0$ \\
Max. & $0.810$ & $1.490$ & $17.0$ \\
\hline
\end{tabular}
\end{table}

\clearpage

\section{Additional Figure in real data application}\label{secA2}

\begin{figure}[h]
\begin{center}
\includegraphics[scale=0.59]{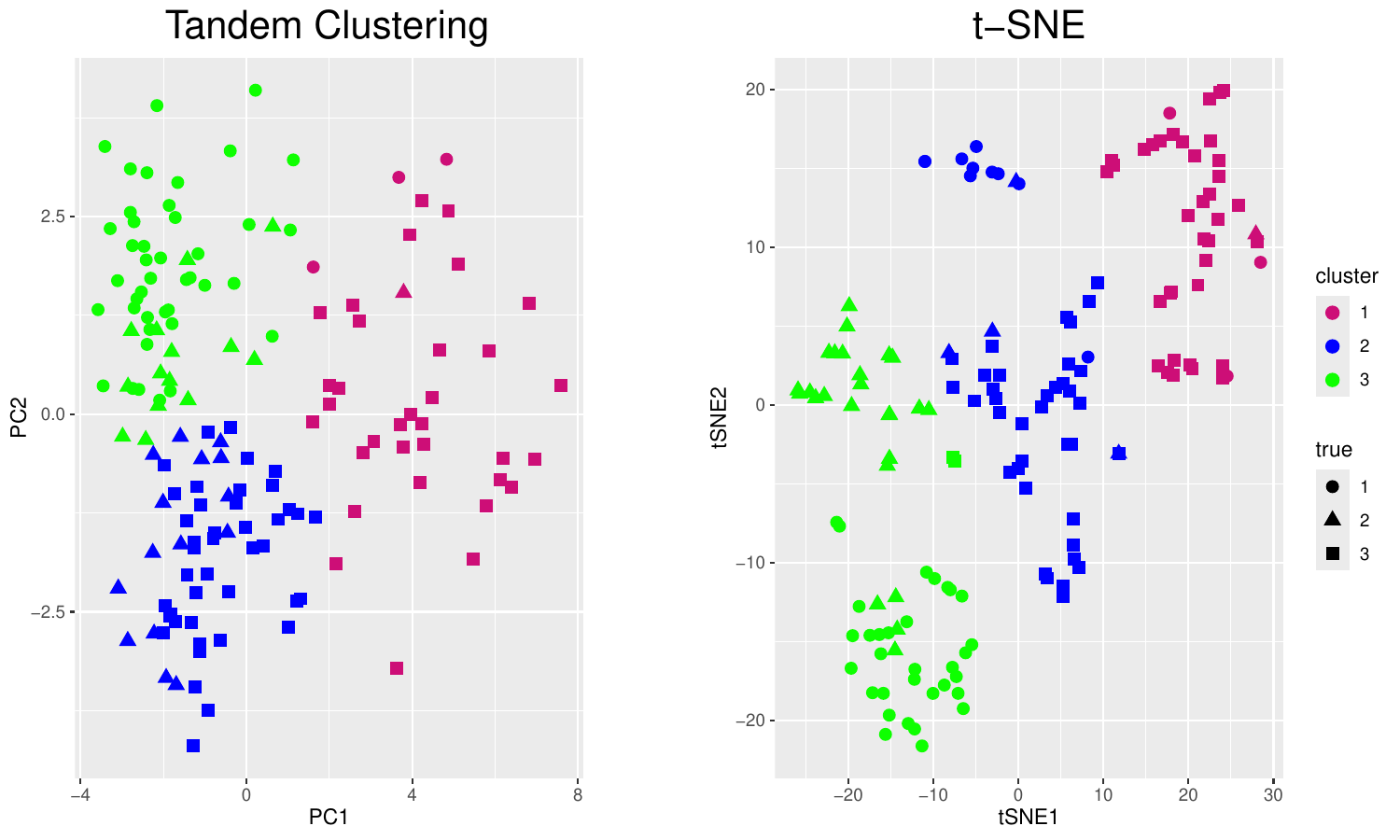}
\caption{Plots of tandem clustering and t-SNE. In tandem clustering, the horizontal axis is the component $1$, and the vertical axis is the component $2$. In t-SNE, the horizontal axis is the first column of the embedding matrix, and the vertical axis is its second column.}
\label{plot_rel_tand_tsne}
\end{center}
\end{figure}

\clearpage

\end{document}